\documentclass[%
reprint,
superscriptaddress,
nofootinbib,
amsmath,
amssymb,
aps,
prd,
floatfix,
showkeys,
]{revtex4-2}

\usepackage[normalem]{ulem}
\usepackage{fancyvrb}
\usepackage{fancyhdr}
\usepackage{graphicx,amsfonts,amssymb,amsbsy}
\usepackage{amsmath,amsthm,latexsym}
\usepackage[utf8]{inputenc}  

\DeclareMathOperator\arcsinh{arcsinh}

\usepackage{natbib}
\usepackage[colorlinks]{hyperref}
\hypersetup{linkcolor=magenta,citecolor=blue}

\begin{document}

\title{Excluded volume effects in the quark-mass density-dependent model: \\ implications for the equation of state and compact star structure}

\author{G. Lugones}
\email{german.lugones@ufabc.edu.br}
\affiliation{Universidade Federal do ABC, Centro de Ci\^encias Naturais e Humanas, Avenida dos Estados 5001- Bang\'u, CEP 09210-580, Santo Andr\'e, SP, Brazil.}

\author{A. G. Grunfeld}
\email{grunfeld@tandar.cnea.gov.ar}
\affiliation{CONICET, Godoy Cruz 2290, Buenos Aires, Argentina} 
\affiliation{Departamento de F\'\i sica, Comisi\'on Nacional de Energ\'{\i}a At\'omica, Avenida del Libertador 8250, (1429) Buenos Aires, Argentina}

\begin{abstract}
We present a significant extension of the quark mass density-dependent model (QMDDM), initially revised in our prior study \cite{Lugones:2022upj}, where thermodynamic inconsistencies were addressed. Our current work enriches the QMDDM by incorporating excluded volume effects, as a step towards a more realistic representation of the quark matter equation of state (EOS) at zero temperature. We introduce the concept of ``available volume'' in the Helmholtz free energy formulation, accounting for the space excluded by each quasiparticle due to its finite size or repulsive interactions. We present a methodology to modify the EOS for point-like particles, allowing for a simple and direct incorporation of excluded volume effects. This is first addressed in a simple one-flavor model and then extended to a more realistic three-flavor system, incorporating both mass and volume dependencies on the baryon number density. We examine various ansatzes for the excluded volume, ultimately adopting one that aligns with the asymptotic freedom behavior of Quantum Chromodynamics (QCD). The EOS for electrically neutral systems in chemical equilibrium is computed, focusing on self-bound and hybrid matter scenarios. We show that the incorporation of excluded volume effects renders the EOS stiffer and that excluded volume effects are essential to align the mass-radius relation of self-bound and hybrid stars with modern astrophysical constraints.

\end{abstract}

\maketitle

\section{Introduction}

In recent years, the study of quark matter at extreme densities has attracted significant attention within the field of nuclear physics and astrophysics, primarily due to its implications in understanding the properties of neutron stars and heavy ion collisions. At the core of these investigations is the EOS, which provides a crucial link between microscopic theories of strong interactions and macroscopic observables. Unfortunately, first-principle calculations for deconfined quark matter are unavailable at present in the high-density, low-temperature conditions expected within the interiors of neutron stars (NS). Consequently, while perturbative QCD imposes certain indirect constraints on the EOS at NS densities \cite{Annala:2019puf}, most insights into quark matter in the NS regime rely on phenomenological models. These models, drawing inspiration from QCD, incorporate key properties of quarks, such as color confinement, asymptotic freedom, and chiral symmetry breaking/restoration, into the EOS. Within this context, the QMDDM has been a topic of discussion in the literature for decades (see e.g. \cite{Fowler:1981rp, Chakrabarty:1989bq, Chakrabarty:1991ui, Chakrabarty:1993db, Benvenuto:1989kc, Lugones:1995vg, Benvenuto:1998tx, Lugones:2002vd, Peng:1999gh, Wang:2000dc, Peng:2000ff, Yin:2008me, Xia:2014zaa} and references therein). In the baseline version of the model,  the system is conceptualized as a noninteracting gas of quasiparticles, each characterized by a mass that varies depending on the baryon number density, $n_B$. This approach hypothesizes that key features of the non-perturbative regime of QCD can be effectively encapsulated through these density-dependent variations in quark masses. 

One aspect of the model that has caused considerable debate in the literature concerns its thermodynamic consistency. Over the years, the model has undergone several reformulations, yet its thermodynamic consistency has never been satisfactorily achieved. However, in a recent article \cite{Lugones:2022upj}, we have demonstrated that the QMDDM can be constructed with thermodynamic consistency when formulated within the canonical ensemble instead of the grand canonical ensemble.

The model presented in Ref. \cite{Lugones:2022upj} is a baseline formulation focusing solely on the variation of quark masses with the density. However, it is necessary to incorporate several other aspects for the model to realistically represent known features of cold and dense matter, such as its remarkably stiffness, as suggested by the observation of very high-mass compact stars  \cite{Demorest:2010bx, Antoniadis:2013pzd, NANOGrav:2019jur, Riley:2021pdl, Miller:2021qha}. This feature is related to the potential existence of repulsive interactions in quark matter that would render the EOS stiffer.
One way to account for this is by considering the excluded volume of particles, akin to the Van der Waals equation of state \cite{Rischke:1991ke, Yen:1997rv, Rischke:1991ke, Steinheimer:2010ib}. The concept of excluded volume arises when particles in close proximity effectively ``exclude" a certain volume around them due to their finite size or repulsive interactions with each other. This reduction in available space for other particles leads to deviations from the ideal behavior predicted by models assuming point-like particles.

In this study, we introduce the effect of excluded volume into the QMDDM. To ensure thermodynamic consistency, we incorporate the excluded volume within the framework of the canonical ensemble. As our focus is on cold compact stellar objects, we proceed with our analysis at zero temperature ($T=0$). 
This paper is structured as follows: In Sec. \ref{sec:generic_1fl}, we develop a methodology for incorporating the effects of excluded volume into a zero-temperature EOS, initially formulated for point-like particles. We focus on a system consisting of a single particle species, with the key assumption that the excluded volume around each particle, is quantified by a function $b(n)$, where $n$ is the particle number density. We show how the standard equations for energy density, pressure, and chemical potential, applicable to point-like particles, can be adapted to include volume exclusion effects. For a quick reference, the summary of our findings in this context can be found in Sec. \ref{sec:sumary_1fl}.
In Sec. \ref{sec:1fl_QMDDM}, we apply the insights from the previous section to reformulate the 1-flavor QMDDM within the canonical ensemble, incorporating excluded volume effects using various ansatzes for $b = b(n)$. This section is primarily pedagogical, aimed at identifying which prescriptions for $b(n)$ are physically viable.
Sec. \ref{sec:generic_3fl} extends the formalism from Sec. \ref{sec:generic_1fl} to a generic EOS involving multiple particle species. The method for incorporating excluded volume effects into an EOS for a mixture of point-like particles is summarized in Sec. \ref{sec:summary_generic_3fl}.
In Sec. \ref{sec:3fl_QMDDM}, we expand the 3-flavor QMDDM, as presented in Ref. \cite{Lugones:2022upj}, to take into consideration excluded volume effects, assuming a generic formula for the excluded volume $b = b(n_B)$.
Finally, in Sec. \ref{sec:3fl_numerical}, we present our numerical results using an ansatz that is consistent  with the asymptotic freedom behavior of QCD. We calculate the EOS under the assumption that matter is electrically charge neutral and in equilibrium under weak interactions, for two different choices of the EOS' parameters: one representing self-bound matter and the other hybrid matter. The structures of self-bound and hybrid stars are studied using these EOS.

\section{Incorporating excluded volume effects in a generic 1-flavor EOS}
\label{sec:generic_1fl}

In this section, we develop a methodology for incorporating the effects of excluded volume into a zero-temperature EOS, originally formulated for point-like particles. For simplicity, we consider a system comprising a single particle species. Our key assumption is that the excluded volume surrounding each particle, attributed to its finite size or repulsive interactions, is quantified by a function $b(n)$, where $n$ is the particle number density. We will show that the equations for energy density, pressure, and chemical potential, applicable to point-like particles, can be straightforwardly adapted to encompass the effect of volume exclusion. For those seeking a concise overview, we direct attention to the summary of our findings in Sec. \ref{sec:sumary_1fl}.

\subsection{EOS for point-like particles}

Below, we provide a summary of some standard definitions and results of thermodynamics that will be useful throughout the text. We will examine a system of $N$ particles within a volume $V$ at absolute zero temperature.  Let us assume that without excluded volume effects, the system can be described by a Helmholtz free energy function denoted as $F_\mathrm{pl}(V, N)$, where the subindex ``pl'' refers to point-like particles. As we are considering the case of $T = 0$, the internal energy is equivalent to $F$, and thus, the energy density is identical  to the Helmholtz free energy per unit volume: 
\begin{equation} 
\epsilon_\mathrm{pl} = \frac{F_\mathrm{pl}(V, N)}{V} . 
\label{eq:first}
\end{equation} 
All other thermodynamic quantities of the system of point-like particles are derived directly from $F_\mathrm{pl}(V, N)$. The pressure is given by:  
\begin{eqnarray}
p_\mathrm{pl} (V, N) =  -  \frac{\partial F_\mathrm{pl}(N,V)}{\partial V} \bigg|_{N} ,
\label{eq:ppp1}
\end{eqnarray}
and the chemical potential by:
\begin{align}
\mu_\mathrm{pl}(V, N) = \frac{\partial F_\mathrm{pl}(N, V)}{\partial N} \bigg|_{V}  .
\label{eq:mmm1}
\end{align}

Taking advantage of the homogeneity properties of thermodynamic functions, the aforementioned quantities can be expressed solely as a function of the particle density $n \equiv N/V$. On one hand, the Helmholtz free energy is a first-order homogeneous function of the extensive parameters. This means that if all extensive parameters of a system are scaled by a factor $\alpha$, the free energy of the resulting system will be scaled by the same factor, according to \cite{Callen}:
\begin{equation}
\alpha F(V, N)  =  F(\alpha V, \alpha N).
\label{eq:homogeneity_1st_order}
\end{equation}
By substituting $\alpha = V^{-1}$ into Eq. \eqref{eq:first}, the energy density can be rewritten as:
\begin{equation}
\epsilon_\mathrm{pl}(n) = \frac{F_\mathrm{pl}(V, N)}{V}  =  F_{\mathrm{pl}}\left(\frac{V}{V}, \frac{N}{V}\right) = F_{\mathrm{pl}}(1, n ).
\label{eq:eee1}
\end{equation}
On the other hand, pressure and chemical potential are intensive quantities, meaning they are zero-order homogeneous functions, as expressed by the following relations \cite{Callen}:
\begin{eqnarray}
p(\alpha V, \alpha N) &=& p(V, N), \\
\mu(\alpha V, \alpha N) &=& \mu(V, N).
\end{eqnarray}
Introducing $\alpha = V^{-1}$ in Eqs. \eqref{eq:ppp1} and \eqref{eq:mmm1}, we obtain:
\begin{eqnarray}
p_{\mathrm{pl}}(n) = p_{\mathrm{pl}}\left(1, \tfrac{N}{V}\right) ,  \\
\mu_{\mathrm{pl}}(n) = \mu_{\mathrm{pl}}\left(1, \tfrac{N}{V}\right) .
\end{eqnarray}

\subsection{EOS with excluded volume effects}
\label{sec:1fl_EOS_with_excluded}

To incorporate the effects of excluded volume, one substitutes the system's volume $V$ with the ``available volume'' $\tilde{V}$, defined as:
\begin{equation}
\tilde{V}=V-b(n) N,
\label{eq:def_tilde_V}
\end{equation}
in the Helmholtz free energy $F_\mathrm{pl}(V, N)$ of point-like particles. In the above equation, $b$ represents the volume that would be excluded by each particle if it were treated as a rigid sphere. To maintain generality, we allowed $b$ to depend on the particle's number density $n$.
Once we replace $V$ with $\tilde{V}$, the resulting function $F_\mathrm{pl}(\tilde{V}, N)$ becomes the starting point for the thermodynamic description that incorporates excluded volume effects.

\subsubsection{Excluded volume effects in the energy density}

At $T = 0$, the energy density is equivalent to the Helmholtz free energy per unit volume:
\begin{equation} 
\epsilon = \frac{F_\mathrm{pl}(\tilde{V}, N)}{V} .
\label{eq:first_2}
\end{equation} 
By applying the homogeneity property from Eq. \eqref{eq:homogeneity_1st_order} with $\alpha = \tilde{V}^{-1}$ to Eq. \eqref{eq:first_2}, the energy density can be reformulated as:
\begin{equation} 
\epsilon = \frac{\tilde{V}}{\tilde{V}} \frac{F_\mathrm{pl}(\tilde{V}, N)}{V} = \frac{\tilde{V}}{V} F_{\mathrm{pl}}\left( \frac{\tilde{V}}{\tilde{V}}, \frac{N}{\tilde{V}}\right).
\label{eq:IIA1_eq12}
\end{equation}

We now introduce the \textit{available volume fraction} $q$, defined as:
\begin{equation}
q(n) \equiv \frac{\tilde{V}}{V} = 1 - n b(n),
\label{eq:definition_of_q}
\end{equation}
which depends solely on the postulated ansatz for the excluded volume per particle, $b(n)$. Note that $q$ satisfies $0 < q < 1$ since $0 < \tilde{V} < V$. The ratio $N/\tilde{V}$ in Eq. \eqref{eq:IIA1_eq12} can be rewritten as:
\begin{equation}
\frac{N}{\tilde{V}} = \frac{N}{V} \frac{V}{\tilde{V}} = \frac{n}{q(n)} .
\end{equation}
Thus, Eq. \eqref{eq:IIA1_eq12} can be reformulated as:
\begin{equation} 
\epsilon(n) = q(n) F_{\mathrm{pl}}\left(1, \frac{n}{q(n)}\right), 
\label{eq:IIA1_eq15}
\end{equation}
which is exclusively a function of $n$.

In Eq. \eqref{eq:IIA1_eq15}, we recognize the energy density for point-like particles, as described by Eq. \eqref{eq:eee1}, as a function of $n/q$. Consequently, we can express the energy density accounting for excluded volume effects in terms of $\epsilon_{\mathrm{pl}}$:
\begin{equation} 
\epsilon(n) = q(n) \epsilon_{\mathrm{pl}} \left(\frac{n}{q(n)} \right). 
\label{eq:eee3}
\end{equation} 
This equation provides a clear method to obtain the energy density with excluded volume corrections. The procedure involves initially expressing $\epsilon_{\mathrm{pl}}$ for point-like particles, which is assumed to be known, as a function of the adjusted variable $n/q(n)$. Subsequently, by multiplying this expression by the correction factor $q(n)$, defined in Eq.~\eqref{eq:definition_of_q}, one can straightforwardly derive the energy density with excluded volume effects. It is important to note that $q(n)$ is dependent solely on the selected ansatz for the excluded volume per particle, denoted by $b(n)$.

\subsubsection{Pressure}
\label{sec:pressure_1fl}

As noted earlier, we account for excluded volume effects by substituting \(\tilde{V} = V - b(n) N\) into the Helmholtz free energy formula for point-like particles. With this substitution,  $F$ continues to be a function of $V$ and $N$, but the dependency on volume is now exclusively represented through the auxiliary function $\tilde{V} = \tilde{V}(V, N)$. To determine the pressure of the system, we must compute the following derivative:
\begin{eqnarray}
p = - \frac{\partial F_{\mathrm{pl}}(\tilde{V}, N)}{\partial V} \bigg|_{N}.
\end{eqnarray}
This derivative can be expressed as:
\begin{equation}
p = - \frac{\partial F_{\mathrm{pl}}}{\partial \tilde{V}} \bigg|_N
\frac{\partial \tilde{V}}{\partial V} \bigg|_N - \frac{\partial F_{\mathrm{pl}}}{\partial N} \bigg|_{\tilde{V}, N} \frac{\partial N}{\partial V} \bigg|_N.
\label{eq:expansion}
\end{equation}
Using the fact that $\partial N/\partial V \big|_N = 0$ and defining:
\begin{align}
\delta \equiv \frac{\partial \tilde{V}}{\partial V} \bigg|_N = \frac{\partial (V - b N)}{\partial V} \bigg|_N = 1 + n^{2} \frac{db}{dn},
\label{eq:definition_of_delta}
\end{align}
Eq. \eqref{eq:expansion} takes the following form:
\begin{equation}
p = - \delta(n) \frac{\partial F_{\mathrm{pl}}(\tilde{V}, N)}{\partial \tilde{V}} \bigg|_N.
\label{eq:aux_20}
\end{equation}

The derivative in the previous equation has the same functional form as the derivative in Eq. \eqref{eq:ppp1}, with the only difference being that it is now evaluated at $\tilde{V}$ instead of at $V$. Consequently, we can identify such derivative as being $p_{\mathrm{pl}}(\tilde{V}, N)$, i.e.:
\begin{eqnarray}
p_{\mathrm{pl}}(\tilde{V}, N) = - \frac{\partial F_{\mathrm{pl}}(\tilde{V}, N)}{\partial \tilde{V}} \bigg|_N.
\label{eq:aux_21}
\end{eqnarray}
Replacing Eq. \eqref{eq:aux_21} into Eq. \eqref{eq:aux_20} we find that the pressure for the system with excluded volume effects can be obtained from the pressure of point-like particles as:
\begin{equation}
p  = \delta(n)  p_\mathrm{pl}(\tilde{V},  N)  .
\end{equation}
Pressure is an intensive quantity, which means it is a homogeneous function of degree zero, that is: $p(\alpha V, \alpha N) = p(V, N)$, for arbitrary $\alpha$ \cite{Callen}. By setting $\alpha = \tilde{V}^{-1}$, we get:
\begin{equation}
p_{\mathrm{pl}}(\tilde{V}, N) = p_{\mathrm{pl}}\left(\frac{\tilde{V}}{\tilde{V}}, \frac{N}{\tilde{V}} \right) = p_{\mathrm{pl}}\left(1, \frac{n}{q} \right) ,
\label{eq:intensive_p}
\end{equation}
where we have used the fact that $N/\tilde{V} = (N/V) \times (V/\tilde{V}) = n/q$.

Therefore, the pressure for the system with excluded volume effects can be obtained from the pressure of point-like particles as:
\begin{equation}
p(n) = \delta(n) p_\mathrm{pl}\left(\tfrac{n}{q}\right) .
\label{eq:p_final_1fl}
\end{equation}

Once more, we have established a simple approach to compute the pressure with corrections accounting for excluded volume. This method begins by representing the known expression $p_\mathrm{pl}$ for point-like particles in terms of the variable $n/ q(n)$. Subsequently, by multiplying it with the correction factor $\delta(n)$, we obtain the pressure incorporating the effect of excluded volume. The correction factors $q(n)$ and $\delta(n)$ are exclusively determined by the chosen ansatz $b(n)$ for the excluded volume per particle.

\subsubsection{Chemical potential}
\label{sec:chemical_1fl}

To obtain the chemical potential we start from:
\begin{eqnarray}
\mu =  \frac{\partial F_{\mathrm{pl}}(\tilde{V}, N)}{\partial N} \bigg|_{V}.
\end{eqnarray}
This derivative can be expressed as:
\begin{equation}
\mu = \frac{\partial F_{\mathrm{pl}} (\tilde{V}, N)   }{\partial \tilde{V}} \bigg|_N
\frac{\partial \tilde{V}}{\partial N} \bigg|_V + \frac{\partial F_{\mathrm{pl}} (\tilde{V}, N)  }{\partial N} \bigg|_{\tilde{V}} .
\label{eq:expansion_mu}
\end{equation}

As already shown, the derivative of $F$ appearing in the first term of the previous equation is $- p_{\mathrm{pl}}(\tilde{V}, N)$. On the other hand, the derivative of the second term has the same functional form as the derivative in Eq. \eqref{eq:mmm1}, with the only difference being that it is now evaluated at $\tilde{V}$ instead of at $V$. Consequently,  we can identify it as being $\mu_{\mathrm{pl}}(\tilde{V}, N)$, i.e.:
\begin{eqnarray}
\mu_{\mathrm{pl}}(\tilde{V}, N) =  \frac{\partial F_{\mathrm{pl}} (\tilde{V}, N)  }{\partial N} \bigg|_{\tilde{V}}  .
\end{eqnarray}
Additionally, we define  
\begin{equation}
- \lambda \equiv \frac{\partial \tilde{V}}{\partial N} \bigg|_V = \frac{\partial (V - b N)}{\partial N} \bigg|_V  =   - \left( b  + n \frac{db}{dn} \right) . 
\label{eq:definition_of_lambda}
\end{equation}
Replacing $p_{\mathrm{pl}}$, $\mu_{\mathrm{pl}}$ and $\lambda$  in Eq. \eqref{eq:expansion_mu} we obtain:
\begin{equation}
\mu = \lambda(n) p_{\mathrm{pl}}(\tilde{V}, N)  + \mu_{\mathrm{pl}}(\tilde{V}, N) .
\label{eq:mu_1fl_789}
\end{equation}
The chemical potential is an intensive quantity; thus, using the same reasoning that led to Eq. \eqref{eq:intensive_p} we get:
\begin{equation}
\mu_{\mathrm{pl}}(\tilde{V}, N)  = \mu_{\mathrm{pl}}\left(1, \frac{n}{q} \right) .
\end{equation}
Therefore, Eq. \eqref{eq:mu_1fl_789} reads:
\begin{equation}
\mu(n) = \lambda(n) p_{\mathrm{pl}} \left(\tfrac{n}{q} \right)  + \mu_{\mathrm{pl}}\left(\tfrac{n}{q} \right) .
\label{eq:fila_chemical_1fl}
\end{equation}
Similar to the case of Eq. \eqref{eq:p_final_1fl}, the above expression takes advantage of the already known expressions $\mu_\mathrm{pl}$ and $p_\mathrm{pl}$ for point-like particles as functions of the variable $n/ q(n)$, along with the correction factor  $\lambda(n)$.

\subsubsection{Assessing thermodynamic consistency via Euler's relation verification}
\label{sec:consistency_1fl}

To confirm the thermodynamic consistency of the model incorporating excluded volume effects, we will check if the Euler relation is verified. We begin by assuming that the system of point-like particles is thermodynamically  consistent and satisfies the Euler relation:
\begin{equation}
\epsilon_{\mathrm{pl}}(V,N) = - p_{\mathrm{pl}}(V,N) + \mu_{\mathrm{pl}}(V,N) \frac{N}{V}. 
\end{equation} 
This relation holds for any system volume, including the volume $\tilde{V}$. Consequently, we obtain:
\begin{equation}
\epsilon_{\mathrm{pl}}(\tilde{V},N) = - p_{\mathrm{pl}}(\tilde{V},N) + \mu_{\mathrm{pl}}(\tilde{V},N) \frac{N}{\tilde{V}},
\end{equation}
which, upon reformulation, yields:
\begin{equation}
\epsilon_{\mathrm{pl}}(\tilde{V},N) = -  p_{\mathrm{pl}}(\tilde{V},N) + \mu_{\mathrm{pl}}(\tilde{V},N) \frac{n}{q} . 
\end{equation}
The preceding equation can be reformulated as follows. First, each term is multiplied by $q = 1 - bn$. Then, we add and subtract the term $p_{\mathrm{pl}} n^2 \frac{db}{dn}$. Finally, the terms are rearranged, leading to the equation below:
\begin{equation} 
q \epsilon_{\mathrm{pl}} = -p_{\mathrm{pl}} - p_{\mathrm{pl}} n^2 \frac{db}{dn} + n \mu_{\mathrm{pl}} + p_{\mathrm{pl}} \left(n \frac{db}{dn} + b\right) n,
\end{equation} 
which simplifies to:
\begin{equation} 
q \epsilon_{\mathrm{pl}} = - \left(1 + n^2 \frac{db}{dn} \right) p_{\mathrm{pl}} + n \left[ \mu_{\mathrm{pl}} + p_{\mathrm{pl}} \left(n \frac{db}{dn} + b \right) \right].
\end{equation} 
This expression incorporates the previously defined factors $\delta$ and $\lambda$, along with the expressions presented in Eqs. \eqref{eq:eee3}, \eqref{eq:p_final_1fl}, and \eqref{eq:fila_chemical_1fl}, culminating in the result:
\begin{equation} 
\epsilon(\tilde{V},N) = - p(\tilde{V},N) + \mu(\tilde{V},N) n.
\end{equation} 
In summary, if the Euler relation is valid for a system of point-like particles, it is also valid for the system with excluded volume effects, thus confirming the thermodynamic consistency of the formalism.

\subsection{Summary of Sec. \ref{sec:generic_1fl}}
\label{sec:sumary_1fl}

This section presented a methodology for taking into account the effects of excluded volume into a pre-existing zero-temperature EOS, initially designed for point-like particles, within the framework of the Helmholtz representation.  Beginning with the established formulas for energy density ($\epsilon_{\mathrm{pl}}$), pressure ($p_{\mathrm{pl}}$), and chemical potential ($\mu_\mathrm{pl}$) of point-like particles, which are functions of the particle number density ($n$), the process unfolds as follows. One first reformulates $\epsilon_{\mathrm{pl}}$, $p_{\mathrm{pl}}$, and $\mu_\mathrm{pl}$ by replacing $n$ with the modified variable $n/q$, where $q$ is defined by the expression given in Eq. \eqref{eq:definition_of_q}:
\begin{equation}
q(n) = 1 - n b(n).
\label{eq:summary_q_1fl}
\end{equation}
In this framework, $b(n)$ symbolizes the per-particle excluded volume, introduced into our model as  a phenomenological ansatz. Then, we derive the correction factors $\delta$ and $\lambda$, which stem from the specific function assigned to $b(n)$, and were defined in Eqs. \eqref{eq:definition_of_delta} and \eqref{eq:definition_of_lambda}, respectively:
\begin{eqnarray}
\delta(n) &=& 1 + n^2 \frac{db}{dn}, \label{eq:summary_delta_1fl}  \\
\lambda(n) &=& b + n \frac{db}{dn}.  \label{eq:summary_lambda_1fl}
\end{eqnarray}
Finally,  the new expressions for energy density, pressure, and chemical potential, now accounting for the effects of excluded volume, are:
\begin{eqnarray}
\epsilon(n) &=& q(n) \times \epsilon_{\mathrm{pl}}\left(\frac{n}{q}\right), \label{eq:summary_epsilon_1fl}   \\
p(n) &=& \delta(n) \times p_{\mathrm{pl}}\left(\frac{n}{q}\right),          \label{eq:summary_p_1fl}   \\
\mu(n) &=& \lambda(n) \times p_{\mathrm{pl}}\left(\frac{n}{q}\right) + \mu_{\mathrm{pl}}\left(\frac{n}{q}\right).  \label{eq:summary_mu_1fl}
\end{eqnarray}
as shown in Eqs. \eqref{eq:eee3}, \eqref{eq:p_final_1fl} and \eqref{eq:fila_chemical_1fl}.

\section{1-Flavor QMDDM: the impact of excluded volume effects}
\label{sec:1fl_QMDDM}

We will now present a reformulation of the QMDDM at $T=0$ focusing on a system with a single particle species. We will account for excluded volume effects by employing various ansatzes for $b = b(n)$. As discussed in the previous section, we will utilize the thermodynamic expressions from the standard QMDDM \cite{Lugones:2022upj}, evaluate them as a function of $n/q$, and incorporate the correction factors $q$, $\delta$, and $\lambda$. For reference, Appendix \ref{sec:appendix_A} summarizes the expressions for the one-flavor QMDDM for point-like particles as derived in Ref. \cite{Lugones:2022upj}.

In order to calculate the energy density, we start by using  Eq. \eqref{eq:appendix_epsilon} in terms of the variable $n/q(n)$. To simplify the notation we introduce an auxiliary variable $\tilde{n}$ defined as:
\begin{equation}
\tilde{n} \equiv \frac{n}{q}.
\end{equation}
In terms of this new variable we have
\begin{equation}
\epsilon_\mathrm{pl}(\tilde{n})  = g  \tilde{M}^4 \chi(\tilde{x}),
\end{equation}
where the function $\chi$ is given in Eq. \eqref{eq:appendix_chi}, $g$ is the particle's degeneracy,  and
\begin{eqnarray}
\tilde{M}    & \equiv &    M(\tilde{n}) = m + \frac{C}{\tilde{n}^{a/3}} ,     \label{eq:mass_with_ecluded_volume}   \\
\tilde{x} & \equiv  &  x(\tilde{n})  = \frac{1}{\tilde{M}} \left( \frac{6 \pi^2 \tilde{n}}{g} \right)^{1/3} .
\end{eqnarray}
In these equations, $m$ represents the current mass, while the constants $a$ and $C$ are treated as free parameters (see Ref. \cite{Lugones:2022upj} for details).

Replacing $\epsilon_\mathrm{pl}(\tilde{n})$ in Eq. \eqref{eq:summary_epsilon_1fl} we obtain the final expression for the energy density with excluded volume effects:
\begin{equation}
\epsilon(n)  = q(n) g  \tilde{M}^4 \chi(\tilde{x}) .
\label{eq:complete_epsilon} 
\end{equation}

To determine the pressure,  we employ Eq. \eqref{eq:appendix_p2} as a function of $\tilde{n}$:
\begin{equation}
p_\mathrm{pl}(\tilde{n})  =  p^{\mathrm{FG}}(\tilde{n}) - B(\tilde{n}) ,
\end{equation}
being:
\begin{eqnarray} 
p^{\mathrm{FG}}(\tilde{n}) & \equiv & g \tilde{M}^{4} \phi(\tilde{x}) , \\ 
B(\tilde{n}) & \equiv &  -g \tilde{M}^{3} \tilde{n} \frac{\partial \tilde{M} }{\partial \tilde{n} } \beta(\tilde{x})>0, \\
\frac{\partial \tilde{M} }{\partial \tilde{n} } & =  & -\frac{C}{3} \frac{a}{\tilde{n}^{a / 3+1}} . 
\end{eqnarray}
The functions $\phi$ and $\beta$ are provided in Eqs. \eqref{eq:appendix_phi} and  \eqref{eq:appendix_beta} respectively. Substituting $p_\mathrm{pl}(\tilde{n})$ into Eq. \eqref{eq:summary_p_1fl}, we obtain the expression for the pressure that accounts for excluded volume effects:
\begin{equation}
p(n) =  \delta(n)  \left[  p^{\mathrm{FG}}(\tilde{n})-B(\tilde{n}) \right]. 
\label{eq:complete_p} 
\end{equation}

Finally,  for the calculation of the chemical potential, we begin with Eq. \eqref{eq:appendix_mu2} expressed in terms of $\tilde{n}$:
\begin{equation}
\mu_\mathrm{pl}(\tilde{n})  = \mu^{\mathrm{FG}}(\tilde{n})-\frac{B(\tilde{n})}{\tilde{n}},
\end{equation}
where $\mu^{\mathrm{FG}}(\tilde{n}) \equiv   \tilde{M} \sqrt{\tilde{x}^{2}+1}$. 
The chemical potential with excluded volume effects is obtained by substituting $\mu_\mathrm{pl}(\tilde{n})$ in Eq. \eqref{eq:summary_mu_1fl}, and reads:
\begin{equation}
\mu(n) = \lambda(n)   \left[   p^{\mathrm{FG}}(\tilde{n}) - B(\tilde{n})  \right]    + \left[   \mu^{\mathrm{FG}}(\tilde{n})-\frac{B(\tilde{n})}{\tilde{n}}  \right] .
\label{eq:complete_mu}
\end{equation}

In  the following subsections we focus on different models for $b = b(n)$.

\subsection{Constant excluded volume}     \label{sec:1fl_constant_b}

The simplest expression for $b = b(n)$ is to take it as a constant. Although this choice is in contradiction with the asymptotic freedom behavior of QCD, because volume exclusion does not vanish as $n \rightarrow \infty$, we will consider it in the following as a starting point.

From Eqs. \eqref{eq:summary_q_1fl}, \eqref{eq:summary_delta_1fl} and \eqref{eq:summary_lambda_1fl},  the functions $q(n)$, $\delta(n)$ and $\lambda(n)$ read:
\begin{eqnarray}
q &=&   1 - n b , \\
\delta &=&   1, \\
\lambda &= & b .
\end{eqnarray}
From Eqs. \eqref{eq:complete_epsilon},  \eqref{eq:complete_p}  and \eqref{eq:complete_mu}  we find: 
\begin{eqnarray}
\epsilon(n)  & = & (1- n b) g  \tilde{M}^4 \chi(\tilde{x}) ,  \label{eq:epsilon_constant_b} \\
p(n) & = &     p^{\mathrm{FG}}(\tilde{n})-B(\tilde{n})  ,  \label{eq:pressure_constant_b} \\
\mu(n) & = &  b  \left[   p^{\mathrm{FG}}(\tilde{n}) - B(\tilde{n})  \right]    + \left[   \mu^{\mathrm{FG}}(\tilde{n})-\frac{B(\tilde{n})}{\tilde{n}}  \right]  , \quad
\label{eq:constantv0_final_mu}
\end{eqnarray}
with     
\begin{equation}
\tilde{n}(n) = \frac{n}{1 - n b}.
\label{eq:n_tilde_fixed_b_1fl}
\end{equation}
These new expressions contain excluded volume corrections through $\tilde{n}$. The chemical potential is additionally corrected by an extra term.

\begin{figure}[tb]
\centering 
 \includegraphics[width=0.97\columnwidth]{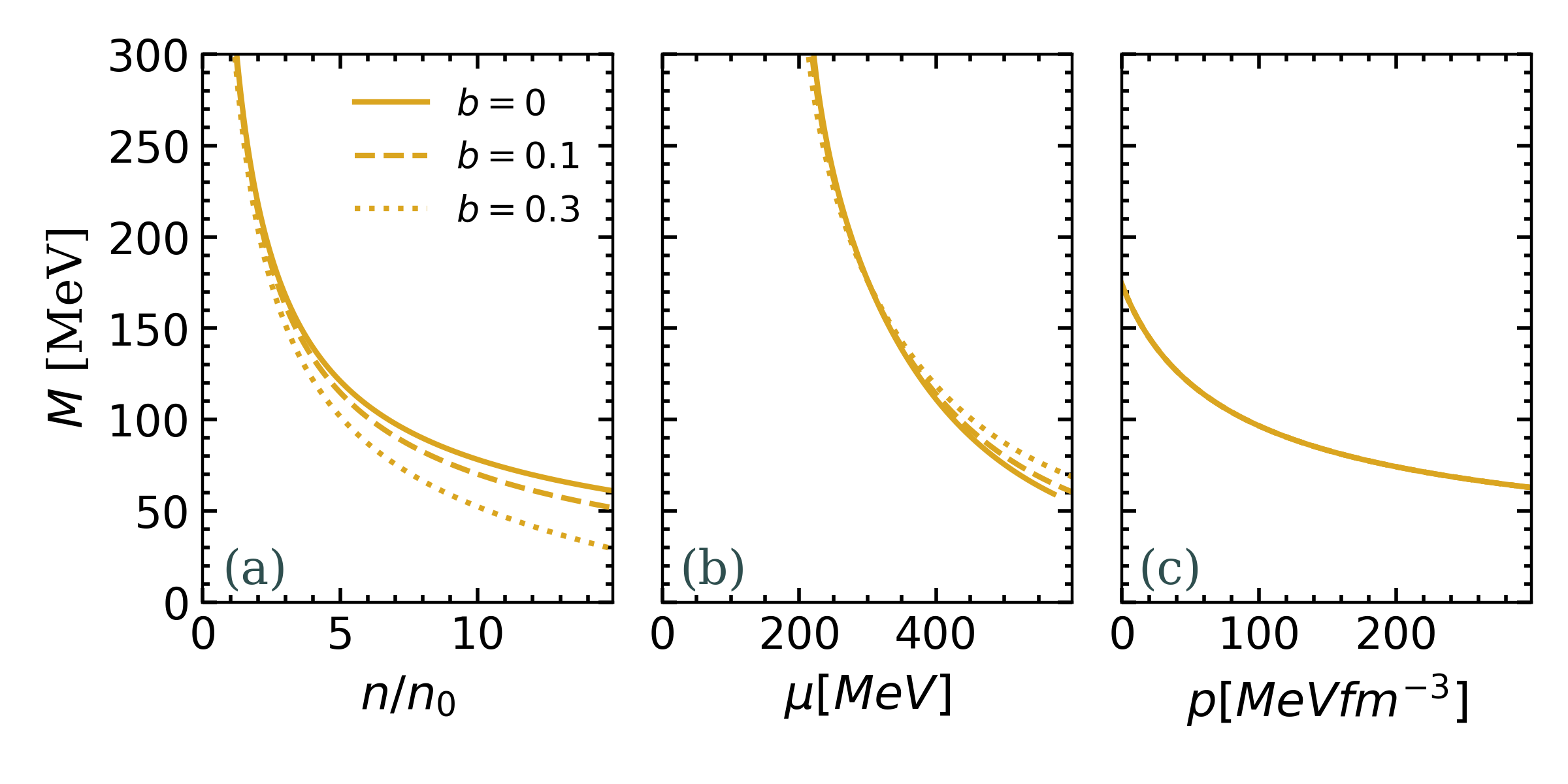}
\caption{Constant excluded volume ansatz: Effective quasiparticle mass $M$ depicted as a function of (a) particle number density $n/n_0$, (b) chemical potential, and (c) pressure. Parameters used are $m=5 ~ \mathrm{MeV}$, $a=2$, and $C=100$, with units assigned such that $n$ is in $\mathrm{fm}^{-3}$ and $M$ is in $\mathrm{MeV}$. The parameter $b$ is in $\mathrm{fm^3}$.  }
\label{fig:constantv0_mass_1}
\end{figure}

In Fig. \ref{fig:constantv0_mass_1} we show the effective quasiparticle mass $M$ for different values of $b$. Note that while the mass formula predicts $M$ to diverge as $n \rightarrow 0$, in practice $M$ has a maximum finite value at zero pressure and approaches the current quark mass $m$ at large pressures and/or densities. Excluded volume effects decrease the effective mass at a constant density, but at a specific pressure, $M$ remains constant for all values of $b$.  This overlapping is due to the fact that both $M$ and $p$ depend directly on $\tilde{n}$ without any explicit dependence on the constant $b$ (cf. Eqs. \eqref{eq:mass_with_ecluded_volume} and \eqref{eq:pressure_constant_b}).

\begin{figure}[tb]
\centering 
 \includegraphics[width=0.97\columnwidth]{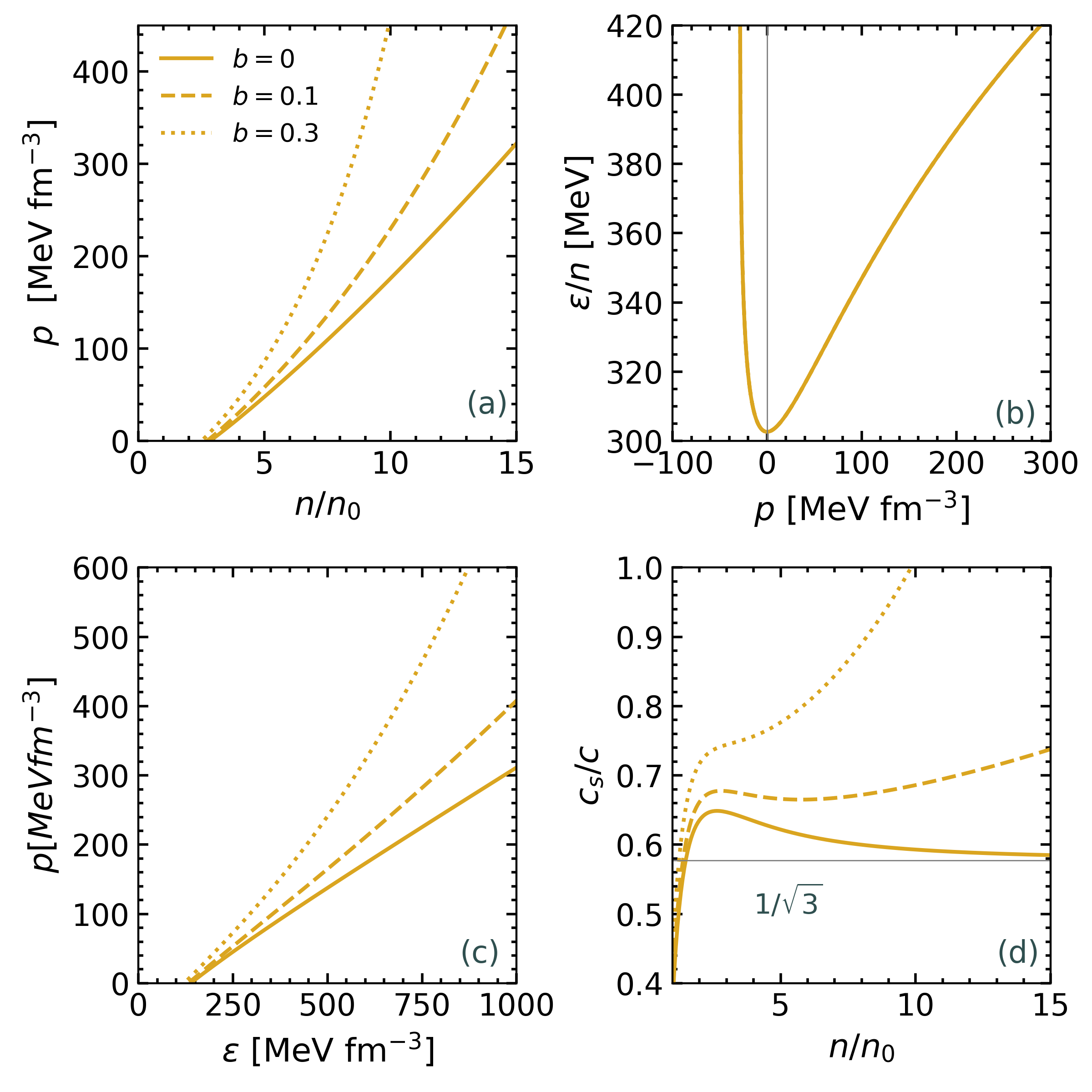}
\caption{EOS for a one-component gas employing the constant excluded volume ansatz, with the same parameter choices as in the preceding  figure. The panels illustrate the following: (a) total pressure $p$ plotted against particle number density $n$ in units of the nuclear saturation density $n_0$; (b) energy per particle $\epsilon/n$ as a function of $p$; (c) pressure versus energy density; and (d) speed of sound as a function of $n/n_0$.}
\label{fig:constantv0_EOS_1}
\end{figure}

In Fig. \ref{fig:constantv0_EOS_1} we show the EOS of a one-component gas for different values of $b$. 
Increasing values of the parameter $b$ result in a stiffer EOS. Note that the curves tend to overlap at zero pressure and clearly separate as the particle density or energy density increases. This behavior arises from the fact that the excluded volume is constant; this means that the role of the excluded volume becomes proportionally less significant when the particles are widely separated (low density).
The minimum of $\epsilon/n$ occurs at $p=0$ as required by thermodynamic consistency.  Notably, the curve for $\epsilon/n$ as a function of pressure is independent of $b$. The above mentioned overlapping results from both $\epsilon/n$ and $p$ having a direct dependence with $\tilde{n}$, without any explicit dependency on $b$ (note that combining Eqs.  \eqref{eq:epsilon_constant_b} and \eqref{eq:n_tilde_fixed_b_1fl}  one finds  $\epsilon/n = g  \tilde{M}^4 \chi(\tilde{x})  / \tilde{n}$).

We also show the speed of sound $c_s$ defined by:
\begin{equation}
c_s^2 = \frac{\partial p}{\partial \epsilon}.
\end{equation}
At densities above a certain threshold, the speed of sound exceeds the speed of light for $b\neq 0$.

Finally, notice that Eq. \eqref{eq:n_tilde_fixed_b_1fl} can be rewritten in terms of the specific volumes $v \equiv 1/n$ and $\tilde{v} \equiv 1/\tilde{n}$ as:
\begin{equation}
\tilde{v} = v - b.
\label{eq:specific_volume_1fl}
\end{equation}
The available volume per particle, $\tilde{v}$, cannot be negative, as that would mean that rigid spheres of volume $b$ ocuppy a volume larger than the system's volume.  Therefore, excluded volume effects are no longer physically meaningful for $v <b$. As a consequence, the EOS is no longer valid at particle number densities exceeding $1/b$. For $b = 0.1 ~ \mathrm{fm^3}$, this threshold density is around $62 n_0$, while for $b = 0.3 ~ \mathrm{fm^3}$ it is approximately $20 n_0$ (being $n_0$ the nuclear saturation density). Moreover, it must be noticed that the causality condition $c_s < c$ is violated at much smaller densities, as seen in Fig. \ref{fig:constantv0_EOS_1}(d). Additionally, $c_s$ does not approach the conformal limit of $1/\sqrt{3}$  at asymptotically high densities.

The above discussion shows that the ansatz with constant $b$ is not satisfactory.  In the following subsection we will explore other formulas for $b(n)$ that fulfill causality and the asymptotic freedom condition of QCD. 

\subsection{Density-dependent excluded volume}

Now, we adopt an ansatz for $b$ in agreement with the asymptotic freedom behavior of QCD, i.e. a formula that allows $b$ to vanish as $n \rightarrow \infty$. 
To this end, $b$ will be expressed as: 
\begin{equation}
b = \kappa n^{-\ell}.
\label{eq:anzatz_decaying_b}
\end{equation}
where $\kappa$ and $\ell$ are positive constants. Note that while this formula predicts $b$ to diverge as $n \rightarrow 0$, in practice $b$ has a maximum finite value at zero pressure.

The values of the parameters $\kappa$ and $\ell$ are in fact interconnected. At  $n \sim n_0$, we expect that repulsive interactions will have a range smaller than about $L \sim 1$ fm, leading to an excluded volume not larger than approximately $\tfrac{4}{3} \pi L^3 \sim 4 \mathrm{fm}^3$. Consequently, if we significantly increase the parameter $\ell$, the constant $\kappa$ cannot be excessively large, as that would result in an overly extensive excluded region.
Taking $\ell=1$ as a reference, Eq. \eqref{eq:anzatz_decaying_b} gives $\kappa = b n$, yielding a maximum value of $\kappa \approx 0.64$ for $n= 0.16 \mathrm{fm}^{-3}$. 
This upper limit is more stringent that the absolute limiting value for $\kappa$ coming from the condition that the available volume cannot be larger than the total volume; i.e. in Eq. \eqref{eq:specific_volume_1fl} $\tilde{v}/v$ must be $< 1$, meaning that for $\ell =1$, $\kappa$ cannot be larger than 1.

In the remainder of this section,  we will adopt the following parameter values: $\kappa = 0,  0.16, 0.64$ together with $\ell=1$. The excluded volume per particle reads:
\begin{equation}
b(n) = \frac{\kappa}{n}, 
\label{eq:66}
\end{equation}
and, using Eqs. \eqref{eq:summary_q_1fl}, \eqref{eq:summary_delta_1fl} and \eqref{eq:summary_lambda_1fl},  the functions $q(n)$, $\delta(n)$ and $\lambda(n)$ are:
\begin{eqnarray}
q &=&   1 - \kappa, \\
\delta &=&  1 - \kappa, \\
\lambda &= & 0 .
\end{eqnarray}
From Eqs. \eqref{eq:complete_epsilon},  \eqref{eq:complete_p}  and \eqref{eq:complete_mu}  we find: 
\begin{eqnarray}
\epsilon(n)  & = & (1- \kappa) g  \tilde{M}^4 \chi(\tilde{x}) , \\
p(n) & = &   (1- \kappa) \left[  p^{\mathrm{FG}}(\tilde{n})-B(\tilde{n}) \right] ,  \\
\mu(n) & = &  \mu^{\mathrm{FG}}(\tilde{n})  -\frac{B(\tilde{n})}{\tilde{n}}  \label{eq:mu_variable_b},
\end{eqnarray}
with
\begin{equation}
\tilde{n}(n) = \frac{n}{1 - \kappa}.
\label{eq:n_tilde_fixed_b}
\end{equation}
Notice that, the functional forms of the pressure and energy density resemble those for point-like particles. The main differences are the introduction of an overall correction factor $1-\kappa$ and the evaluation of the functions at  $\tilde{n}$  rather than $n$. The expression for the chemical potential remains the same as for point-like particles, with the sole distinction being that it is now evaluated at $\tilde{n}$.

\begin{figure}[tb]
\centering 
  \includegraphics[width=0.97\columnwidth]{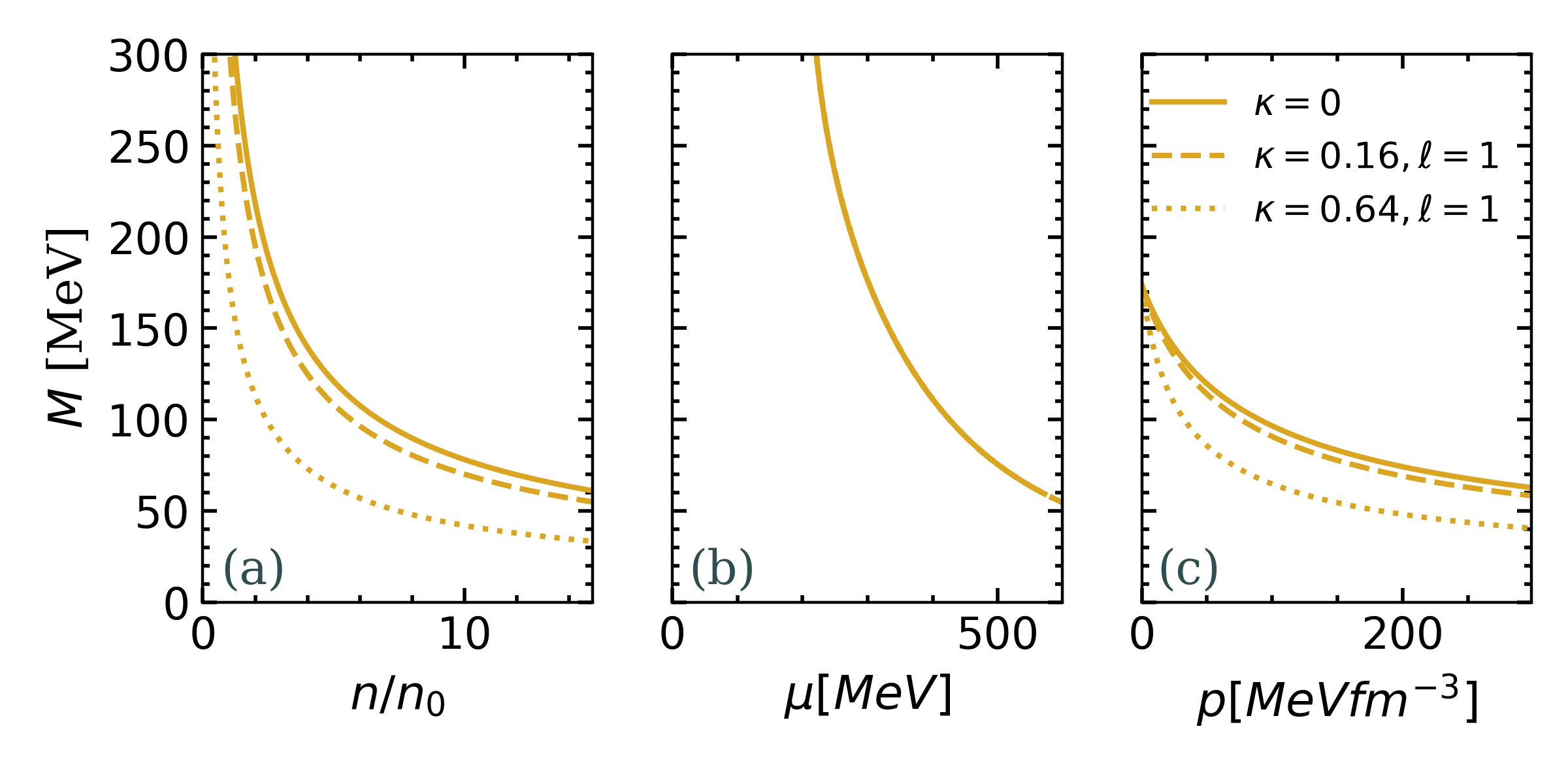}
\caption{Effective quasiparticle mass $M$ in the density-dependent excluded volume ansatz. Panels follow the organization of Fig. \ref{fig:constantv0_mass_1}, with identical choices for the parameters $m$, $a$ and $C$.}
\label{fig:variable_b_mass}
\end{figure}

In Fig. \ref{fig:variable_b_mass}, we display the effective quasiparticle mass $M$ as a function of  density, chemical potential, and pressure. We considered different $\kappa$ values with a constant exponent $\ell = 1$. 
The mass $M$ exhibits the same qualitative behavior as that of point-like particles; that is, it has a finite maximum value at zero pressure and gradually approaches the current mass $m$ at extremely high pressures or densities.
For specific density or pressure values, $M$ is always lower for higher $\kappa$. Interestingly, the mass curves for different $\kappa$ values match when plotted against the chemical potential. This is because both $M$ and $\mu$ depend directly on $\tilde{n}$ without any explicit dependence on $\kappa$, as shown in Eqs. \eqref{eq:mass_with_ecluded_volume} and \eqref{eq:mu_variable_b}.

\begin{figure}[tbh]
\centering 
  \includegraphics[width=0.97\columnwidth]{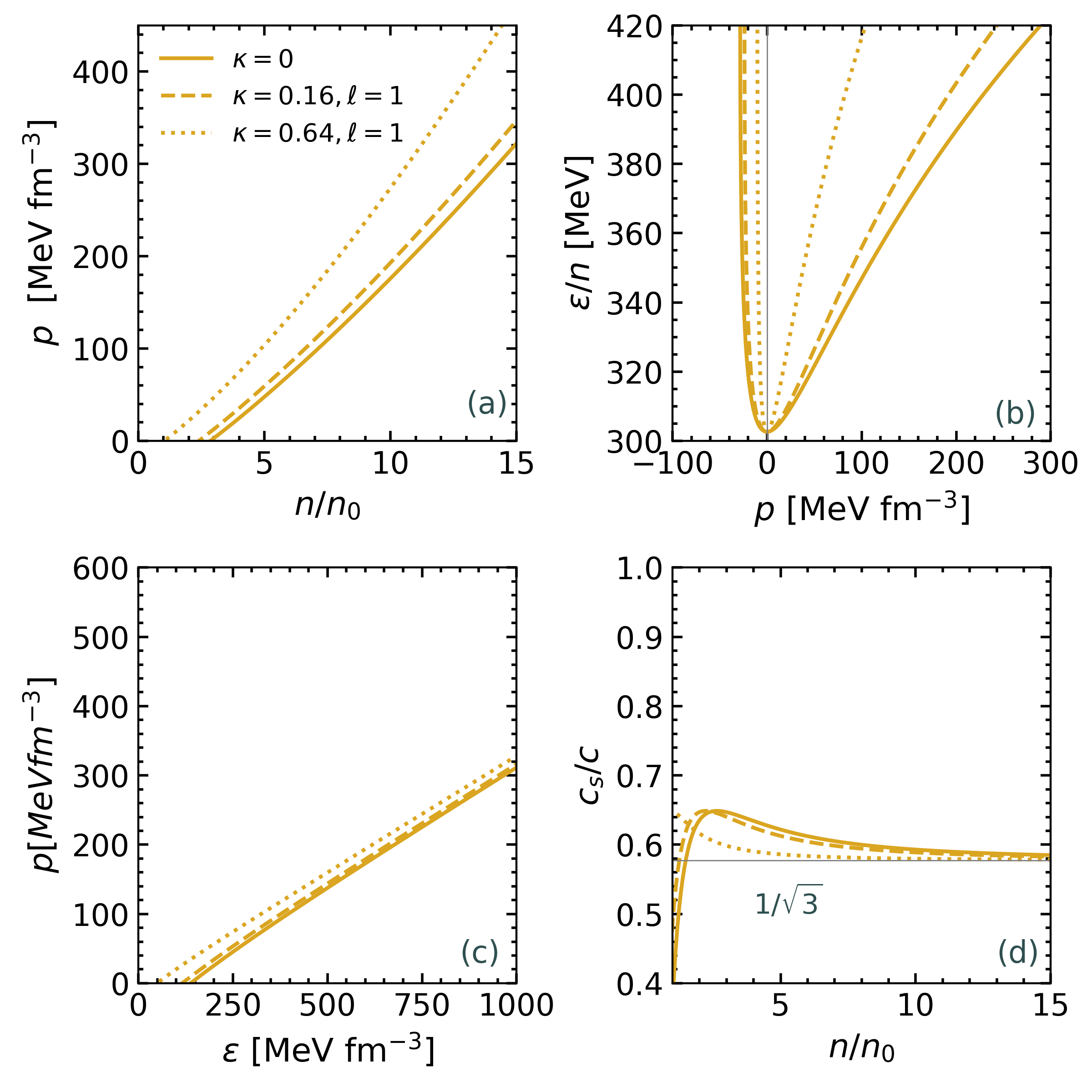}
\caption{EOS for a one-component gas with the density-dependent excluded volume ansatz. Panels are organized as in Fig. \ref{fig:constantv0_EOS_1}, with identical choices for the parameters $m$, $a$ and $C$. }  
\label{fig:variable_b_EOS}
\end{figure}

In Fig. \ref{fig:variable_b_EOS}, we present the EOS for various values of $\kappa$. The pressure behavior displayed in Fig. \ref{fig:variable_b_EOS} is qualitatively different from that in Fig. \ref{fig:constantv0_EOS_1}. As seen in panels (a) and (c), the curves for different $\kappa$ values are separated at all densities. The curves do not overlap at low $n$ because the excluded volume is inversely dependent on the density. As a result, the volume exclusion effect becomes more significant at lower densities. As density increases, volume exclusion vanishes, and the separation between the curves tends to remain constant, as observed in panel (c) of Fig. \ref{fig:variable_b_EOS}. For the same reason, all the speed of sound curves converge to the conformal limit of $c_s = 1/\sqrt{3}$ for asymptotically large densities. Another relevant aspect of the EOS is its increasing stiffness as $\kappa$ rises, as seen in panels (a) and (c) of Fig. \ref{fig:variable_b_EOS}. As seen in panel (b) of Fig. \ref{fig:variable_b_EOS}, the minimum of $\epsilon/n$ occurs at $p=0$, as required for thermodynamic consistency. Contrary to the case of Sec. \ref{sec:1fl_constant_b}, the curves no longer overlap.

In summary, in contrast to the constant excluded-volume approach discussed in the previous subsection, the density-dependent ansatz provided by Eq. \eqref{eq:66} yields a viable quark matter EOS, in agreement with qualitative aspects of QCD.

\section{Excluded volume effects in a generic 3-flavor EOS}
\label{sec:generic_3fl}

For applications in astrophysics, the most relevant state of quark matter involves an electrically neutral mixture of up ($u$), down ($d$), and strange ($s$) quarks together with a minor proportion of electrons ($e$), with all components in a state of chemical equilibrium due to the influence of weak interactions. Below, we will expand upon the equations presented in Sec. \ref{sec:generic_1fl}  to accommodate the scenario involving all three quark flavors.

\subsection{EOS for point-like particles}

Assuming the absence of excluded volume effects, let the Helmholtz free energy function be denoted as $F_{\text{pl}}(V, \{N_j\})$, where ``pl'' represents point-like particles. Here, $\{N_j\}$ represents the set $\{N_u, N_d, N_s\}$, with $N_u$, $N_d$, and $N_s$ being the total number of particles for the $u$, $d$, and $s$ quarks, respectively. At $T = 0$, the energy density is identical  to the Helmholtz free energy per unit volume: 
\begin{equation} 
\epsilon_\mathrm{pl} = \frac{F_\mathrm{pl}(V, \{N_j\})}{V} . 
\label{eq:first_3fl}
\end{equation} 
All other thermodynamic quantities of the system of point-like particles are derived directly from $F_\mathrm{pl}(V, \{N_j\})$. The pressure is given by:  
\begin{eqnarray}
p_\mathrm{pl} (V, \{N_j\}) =  -  \frac{\partial F_\mathrm{pl}(V, \{N_j\})}{\partial V} \bigg|_{  \{N_j\}  } ,
\label{eq:ppp1_3fl}
\end{eqnarray}
and the chemical potential by:
\begin{align}
\mu_{\mathrm{pl}, k}(V, \{N_j\}) = \frac{\partial F_\mathrm{pl}(V, \{N_j\})}{\partial N_k} \bigg|_{V,   \{N_{j \neq k}  \}  }  .
\label{eq:mmm1_3fl}
\end{align}

Using the extensivity of $F$ and the intensivity of the pressure and chemical potential  \cite{Callen}, i.e.
\begin{eqnarray}
F_\mathrm{pl}(\alpha V,  \{\alpha N_j\}) & = & \alpha F_\mathrm{pl}(V, \{N_j\}),    \label{eq:homogeneity_1st_order_3fl}   \\
p_{\mathrm{pl}}(\alpha V,  \{\alpha N_j\}) &=& p_{\mathrm{pl}}(V, \{N_j\}), \\
\mu_{\mathrm{pl}, k}(\alpha V,  \{\alpha N_j\}) &=& \mu_{\mathrm{pl}, k}(V, \{N_j\}),
\end{eqnarray}
we can express all thermodynamic quantities in terms of the particle number densities $n_i \equiv N_i/V$. By setting $\alpha = V^{-1}$ the equations transform into:
\begin{eqnarray}
\epsilon_\mathrm{pl}(\{n_j\}) & = & \frac{F_\mathrm{pl}(V, \{N_j\})}{V}  =  F_{\mathrm{pl}}\left(\frac{V}{V},  \left\{ \frac{N_j}{V} \right\}  \right)  \nonumber \\
     & = &  F_{\mathrm{pl}}(1, \{n_j\} ),       \label{eq:eee1_3fl}     \\
p_{\mathrm{pl}}( \{n_j\} ) &=&  p_{\mathrm{pl}}   \left(1, \left\{ \tfrac{N_j}{V} \right\} \right) ,  \\
\mu_{\mathrm{pl}, k}( \{n_j\} ) &=&  \mu_{\mathrm{pl}, k}   \left(1, \left\{ \tfrac{N_j}{V} \right\} \right) .
\end{eqnarray}

\subsection{EOS with excluded volume effects}

To account for excluded volume effects we replace the system's volume $V$ by the available volume $\tilde{V}$. Since volume exclusion is the result of repulsion associated with strong interactions, the natural generalization of the one flavor expression given in Eq. \eqref{eq:def_tilde_V} is the following  flavor independent formula:
\begin{align}
\tilde{V}=V -  b(n_B) N_B, 
\label{eq:volume_nB}
\end{align}
where $N_B$ is the total baryon number of the system and $b$ represents the volume that would be excluded by each particle if it were treated as a rigid sphere. 
To maintain generality, we allow $b$ to depend on the baryon number density $n_B$, which is given by:
\begin{equation} 
n_{B}=\tfrac{1}{3}\left(n_{u}+n_{d}+n_{s}\right) .
\end{equation} 

Once we replace $V$ with $\tilde{V}$ in the Helmholtz free energy $F_\mathrm{pl}(V, \{N_i\})$ of point-like particles, the resulting function $F_\mathrm{pl}(\tilde{V}, \{N_i\})$ becomes the fundamental starting point for the thermodynamic description that incorporates excluded volume effects.

\subsubsection{Energy density}

At $T = 0$, the energy density is equivalent to the Helmholtz free energy per unit volume:
\begin{equation} 
\epsilon = \frac{F_\mathrm{pl}(\tilde{V}, \{N_j\})}{V} .
\label{eq:first_2_3fl}
\end{equation} 
By applying the homogeneity property from Eq. \eqref{eq:homogeneity_1st_order_3fl} with $\alpha = \tilde{V}^{-1}$ to Eq. \eqref{eq:first_2_3fl}, the energy density can be reformulated as:
\begin{equation} 
\epsilon = \frac{\tilde{V}}{\tilde{V}} \frac{F_\mathrm{pl}(\tilde{V}, \{N_j\}  )}{V} = \frac{\tilde{V}}{V} F_{\mathrm{pl}}\left( \frac{\tilde{V}}{\tilde{V}},  \left\{ \frac{N_j}{\tilde{V}} \right\}  \right).
\label{eq:IIA1_eq12_3fl}
\end{equation}

We now introduce the \textit{available volume fraction} $q$, defined as:
\begin{equation}
q(n_B) \equiv \frac{\tilde{V}}{V} =   1- n_B b(n_B) ,
\label{eq:3fl_q_definition}
\end{equation}
which depends solely on the postulated ansatz for the excluded volume per particle, $b(n_B)$. Note that $q$ satisfies $0 < q < 1$ since $0 < \tilde{V} < V$. The ratio $N_j/\tilde{V}$ in Eq. \eqref{eq:IIA1_eq12_3fl} can be rewritten as:
\begin{equation}
\frac{N_j}{\tilde{V}} = \frac{N_j}{V} \frac{V}{\tilde{V}} = \frac{n_j}{q(n_B)}
\end{equation}
Thus, Eq. \eqref{eq:IIA1_eq12_3fl} can be reformulated as:
\begin{equation} 
\epsilon(\{n_j\}) = q(n_B) F_{\mathrm{pl}}\left(1, \left\{ \frac{n_j}{q(n_B)} \right\}  \right), 
\label{eq:IIA1_eq15_3fl}
\end{equation}
which is exclusively a function of $\{n_j\}$.

In Eq. \eqref{eq:IIA1_eq15_3fl}, we identify the energy density for point-like particles, as depicted by Eq. \eqref{eq:eee1_3fl}, expressed as a function of $\{n_j/q\}$. This allows us to obtain the energy density that incorporates excluded volume effects in terms of \(\epsilon_{\mathrm{pl}}\):
\begin{equation} 
\epsilon(\{n_j\}) = q(n_B) \epsilon_{\mathrm{pl}} \left(\left\{ \frac{n_j}{q(n_B)} \right\}  \right). 
\label{eq:eee3_3fl}
\end{equation} 
This equation provides a straightforward procedure to calculate the energy density with excluded volume corrections. The process begins by expressing $\epsilon_{\mathrm{pl}}$ for point-like particles, which we assume as a known quantity, as a function of the modified variables $\{n_j/q(n_B)\}$ and  then multiplying the result by the correction factor $q(n_B)$ defined in Eq. \eqref{eq:3fl_q_definition}. As emphasized earlier, $q(n_B)$ depends solely on the chosen ansatz for the excluded volume per particle $b(n_B)$.

\subsubsection{Pressure}

As discussed before, excluded volume effects are taken into account by substituting $\tilde{V} = V - b(n_B) N_B$ into the Helmholtz free energy for point-like particles. With such replacement, $F$ will still be a function of $V$ and $\{N_j\}$, but the dependence on volume occurs only through the auxiliary function $\tilde{V} = \tilde{V}(V, \{N_j\})$. To determine the pressure of the system we follow the same procedure as in Sec. \ref{sec:pressure_1fl}. We start with the pressure definition:
\begin{eqnarray}
p(V, \{N_j\}) =  -  \frac{\partial F_\mathrm{pl}(\tilde{V}, \{N_j\})}{\partial V} \bigg|_{  \{N_j\}  } ,
\end{eqnarray}
which can be expressed as:
\begin{equation}
\begin{aligned}
\frac{\partial F_\mathrm{pl}}{\partial V}\bigg|_{\{N_{j}\}}
= &  \frac{\partial F_\mathrm{pl}}{\partial \tilde{V} }\bigg|_{\{N_{j}\}}  
\frac{\partial \tilde{V}}{\partial V} \bigg|_{\{N_{j}\}}   \\
& + \sum_{i} \frac{\partial F_\mathrm{pl}}{\partial N_{i}}\bigg|_{  \tilde{V}, \{ N_{j} \neq i \} }  \frac{\partial N_{i}}{\partial V}\bigg|_{\left\{N_{j}\right\}} .
\label{eq:expansion_3fl}
\end{aligned}   
\end{equation}
Using the fact that ${\partial N_{i}} / {\partial V}\big|_{\left\{N_{j}\right\}} = 0$ and defining $\delta$ as:
\begin{align}
\delta \equiv \frac{\partial \tilde{V}}{\partial V} \bigg|_{\{N_{j}\}} = \frac{\partial (V - b N_B)}{\partial V} \bigg|_{\{N_{j}\}} = 1 + n_B^{2} \frac{db}{dn_B},
\label{eq:definition_of_delta_3fl}
\end{align}
we can reformulate Eq. \eqref{eq:expansion_3fl} in the following manner:
\begin{equation}
p = - \delta(n_B) \frac{\partial F_{\mathrm{pl}}(\tilde{V}, \{N_j\}   )}{\partial \tilde{V}} \bigg|_{ \{N_j\} } .
\label{eq:89_3fl}
\end{equation}

The derivative in the previous equation has the same functional form as the derivative in Eq. \eqref{eq:ppp1_3fl}, but it is now evaluated at $\tilde{V}$ instead of at $V$. Therefore, it can be identified as the point-like pressure $-p_{\mathrm{pl}}(\tilde{V}, \{N_j\})$. Consequently, Eq. \eqref{eq:89_3fl} reads:
\begin{equation}
p(V, \{N_j\}) = \delta(n_B) p_{\mathrm{pl}}(\tilde{V}, \{N_j\}).
\end{equation}

Given that pressure is an intensive quantity, it satisfies the condition $p(\alpha V, \alpha N) = p(V, N)$, for arbitrary $\alpha$  \cite{Callen}. Setting $\alpha = \tilde{V}^{-1}$, we obtain:
\begin{equation}
\begin{aligned}
p_{\mathrm{pl}}(\tilde{V}, \{N_j\} ) & = p_{\mathrm{pl}}\left(\frac{\tilde{V}}{\tilde{V}}, \left\{ \frac{N_j}{\tilde{V}}  \right\} \right)  \\
& = p_{\mathrm{pl}}\left(1,   \left\{ \frac{n_j}{q}  \right\}   \right) ,
\label{eq:intensive_p_3fl}
\end{aligned}
\end{equation}
where we have used the fact that $N_j/\tilde{V} = (N_j/V) \times (V/\tilde{V}) = n_j/q$.

Therefore, the pressure for the system with excluded volume effects can be obtained from the pressure of point-like particles as:
\begin{equation}
p( \{n_j\} ) = \delta(n_B) p_\mathrm{pl}\left(  \left\{ \tfrac{n_j}{q}  \right\}   \right) .
\label{eq:p_final_3fl}
\end{equation}
Once again, we arrive at a straightforward procedure for calculating the pressure with excluded volume corrections. We begin with the known expression for the pressure $p_{\mathrm{pl}}$ for point-like particles, rewrite it in terms of the variable set $\{n_j/q\}$, and multiply by the factor $\delta$. Both $q$ and $\delta$ are uniquely determined by the chosen ansatz $b(n_B)$ for the excluded volume per particle.

\subsubsection{Chemical potential}

To obtain the chemical potential we  proceed as in  Sec.  \ref{sec:chemical_1fl}.  We start from:
\begin{eqnarray}
\mu_k(V, \{N_j\}) =  \frac{\partial F_\mathrm{pl}(\tilde{V}, \{N_j\})}{\partial N_k} \bigg|_{V,  \{N_{j \neq k} \}  } ,
\end{eqnarray}
which can be expressed as:
\begin{equation}
\begin{aligned}
\mu_k = &   \frac{\partial F_\mathrm{pl} } {\partial \tilde{V} } \bigg|_{ \{N_ j \} }   \frac{\partial \tilde{V}}{\partial N_k  } \bigg|_{V,  \{ N_{j \neq k} \} }  \\
&  +    \sum_i \frac{\partial F_\mathrm{pl} }  {\partial N_{i}}\bigg|_{\tilde{V}, N_{j} \neq i}  \frac{\partial N_i }{\partial N_k}  \bigg|_{V,  \{ N_{j \neq k} \}  }  .
\end{aligned}
\label{eq:expansion_mu_3fl}
\end{equation}
As already shown, the derivative of $F$ appearing in the first term of the previous equation is $-p_{\mathrm{pl}}(\tilde{V}, \{N_j\})$. On the other hand, the derivative of $F$ in the second term has the same functional form as the derivative in Eq. \eqref{eq:mmm1_3fl}, with the only difference being that it is now evaluated at $\tilde{V}$ instead of at $V$. Consequently,
\begin{align}
\mu_{\mathrm{pl}, k}(\tilde{V}, \{N_j\}) = \frac{\partial F_\mathrm{pl}(\tilde{V}, \{N_j\})}{\partial N_k} \bigg|_{\tilde{V},   \{N_{j \neq k}  \}  }  .
\end{align}
Additionally, we define  
\begin{equation}
\begin{aligned}
- \lambda \equiv & \frac{\partial \tilde{V}}{\partial N_k  } \bigg|_{V,  \{ N_{j \neq k} \} }  = \frac{\partial (V - b N_B)}{\partial N_k} \bigg|_{V,  \{ N_{j \neq k} \} }  \\
= &     -  \tfrac{1}{3} \left( b  + n_B \frac{db}{dn_B} \right) .
\label{eq:definition_of_lambda_3fl}
\end{aligned}
\end{equation}
Replacing the prior expressions in Eq. \eqref{eq:expansion_mu_3fl} we obtain:
\begin{equation}
\mu_k = \lambda(n_B) p_{\mathrm{pl}}(\tilde{V}, \{N_j\}) + \mu_{\mathrm{pl}, k}(\tilde{V}, \{N_j\}) .
\label{eq:mu_3fl_789}
\end{equation}
The chemical potential is an intensive quantity; thus, using the same reasoning that led to Eq. \eqref{eq:intensive_p_3fl} we get:
\begin{equation}
\mu_{\mathrm{pl}, k}(\tilde{V}, \{N_j\})  = \mu_{\mathrm{pl}, k}   \left(1,   \left\{ \frac{n_j}{q}  \right\}   \right)  .
\end{equation}
Therefore, Eq. \eqref{eq:mu_3fl_789} reads:
\begin{equation}
\mu_k( \{n_j\} )  = \lambda(n_B) p_{\mathrm{pl}} \left(  \left\{ \tfrac{n_j}{q}  \right\}   \right)  + \mu_{\mathrm{pl}, k}  \left(  \left\{ \tfrac{n_j}{q}  \right\}   \right) .
\label{eq:fila_chemical_3fl}
\end{equation}
Similar to the case of Eq. \eqref{eq:p_final_3fl}, the above expression takes advantage of the already known expressions $\mu_\mathrm{pl}$ and $p_\mathrm{pl}$ for point-like particles as functions of the variable set $\{n_j/q\}$, along with the correction factor  $\lambda$.

\subsubsection{Thermodynamic consistency: a look at Euler's relation}

To verify the thermodynamic consistency of the formulas incorporating excluded volume effects, we begin by postulating that the system of point-like particles is thermodynamically consistent and satisfies the Euler relation (cf. Sec. \ref{sec:consistency_1fl}):
\begin{equation}
\epsilon_{\mathrm{pl}}(V, \{N_j\}) = - p_{\mathrm{pl}}(V, \{N_j\}) + \sum_k \mu_{\mathrm{pl}, k}(V, \{N_j\}) \frac{N_k}{V}. 
\end{equation}
This relation holds for any system volume, including the volume $\tilde{V}$. Consequently, we obtain:
\begin{equation}
\epsilon_{\mathrm{pl}}(\tilde{V}, \{N_j\})  =  - p_{\mathrm{pl}}(\tilde{V}, \{N_j\}) + \sum_k \mu_{\mathrm{pl}, k}(\tilde{V}, \{N_j\}) \frac{N_k}{\tilde{V}} .
\end{equation}
Using $N_k / \tilde{V} =  n_k / q$ and reorganizing the previous expression one obtains: 
\begin{equation}
 q \epsilon_{\mathrm{pl}}(\tilde{V}, \{N_j\}) 
 = -  q p_{\mathrm{pl}}(\tilde{V}, \{N_j\}) + \sum_k \mu_{\mathrm{pl}, k}(\tilde{V}, \{N_j\}) n_k.  
\end{equation}
We first substitute $q$ with $1 - bn_B$. Next, we add and subtract the term $p_{\mathrm{pl}} n_B^2 \frac{db}{dn_B}$  to the equation. After rearranging the terms, the final form of the equation is:
\begin{equation} 
\begin{aligned}
q \epsilon_{\mathrm{pl}}  = &  -p_{\mathrm{pl}} - p_{\mathrm{pl}} n_B^2 \frac{db}{dn_B} \\
   & + \sum_k n_k \mu_{\mathrm{pl}, k} + p_{\mathrm{pl}} \left(n_B \frac{db}{dn_B} + b \right) n_B,
\end{aligned}
\end{equation} 
which simplifies to:
\begin{equation} 
\begin{aligned}
q \epsilon_{\mathrm{pl}} = & - \left(1 + n_B^2 \frac{db}{dn_B} \right) p_{\mathrm{pl}}  \\ 
& + \sum_k n_k \left[ \mu_{\mathrm{pl}, k}  + p_{\mathrm{pl}} ~ \tfrac{1}{3} \left(n_B \frac{db}{dn_B} + b \right) \right].
\end{aligned}
\end{equation} 
This expression incorporates the previously defined factors $\delta$ and $\lambda$, along with the expressions presented in Eqs. \eqref{eq:eee3_3fl}, \eqref{eq:p_final_3fl}, and \eqref{eq:fila_chemical_3fl}, culminating in the result:
\begin{equation} 
\epsilon(\tilde{V}, \{N_j\} ) = - p(\tilde{V}, \{N_j\} ) + \sum_k \mu_k(\tilde{V}, \{N_j\} ) n_k.
\end{equation} 
In summary, if the Euler relation is valid for a system of point-like particles, it is equally valid for the system with excluded volume effects, thus confirming the thermodynamic consistency of the formalism.

\subsection{Summary of Sec. \ref{sec:generic_3fl}}
\label{sec:summary_generic_3fl}

In this section, we showed how to straightforwardly incorporate excluded volume effects into any zero-temperature EOS already formulated for point-like particles in the Helmholtz representation.  We begin by considering the established expressions for the energy density $\epsilon_{\mathrm{pl}}$, the pressure $p_{\mathrm{pl}}$, and the chemical potentials $\mu_{\mathrm{pl}, i}$ of point-like particles, which depend on the set of particle number densities $\{n_j\}$. Initially, we rewrite $\epsilon_{\mathrm{pl}}$, $p_{\mathrm{pl}}$, and $\mu_{\mathrm{pl}, i}$ using the modified variable set $\{n_j/q\}$, where $q$ is defined as: 
\begin{equation}
q(n_B) =   1- n_B b(n_B) ,
\label{eq:summary_q_3fl}
\end{equation}
in accordance with Eq. \eqref{eq:3fl_q_definition}. In this context, $b(n_B)$ represents the excluded volume per particle and is incorporated into the model as a phenomenological ansatz. 
Next, we determine the correction factors $\delta$ and $\lambda$, which are defined in Eqs. \eqref{eq:definition_of_delta_3fl} and \eqref{eq:definition_of_lambda_3fl}, respectively:
\begin{eqnarray} 
\delta &=&  1 + n_B^{2} \frac{db}{dn_B}, \label{eq:summary_delta_3fl} \\
\lambda  &=&  \tfrac{1}{3} \left( b  + n_B \frac{db}{dn_B} \right) .   \label{eq:summary_lambda_3fl}
\end{eqnarray}
These factors are derived directly from the specified function for $b(n_B)$.
Finally, as shown in Eqs. \eqref{eq:eee3_3fl}, \eqref{eq:p_final_3fl} and \eqref{eq:fila_chemical_3fl}, the energy density, pressure, and chemical potentials,  incorporating excluded volume effects, are expressed as follows:
\begin{eqnarray}
\epsilon(\{n_j \}) & = & q \sum_{i}  \epsilon_{\mathrm{pl},i}    ( \left\{ n_j /q  \right\} )  , \label{eq:summary_epsilon_3fl} \\
p( \left\{ n_j \right\} ) & = &   \delta  \sum_{i}   p_{\mathrm{pl},i} ( \left\{ n_j /q  \right\} ) , \label{eq:summary_p_3fl}\\
\mu_k( \{n_j\} ) & = & \lambda(n_B) p_{\mathrm{pl}} \left(  \left\{ \tfrac{n_j}{q}  \right\}   \right)  + \mu_{\mathrm{pl}, k}  \left(  \left\{ \tfrac{n_j}{q}  \right\}   \right) . \qquad  \label{eq:summary_mu_3fl}
\end{eqnarray}

\section{3-flavor QMDDM with excluded volume effects: FORMALISM}
\label{sec:3fl_QMDDM}

In this section, we use the formalism presented in Sec. \ref{sec:generic_3fl} to incorporate excluded volume effects in the 3-flavor QMDDM EOS developed in Ref. \cite{Lugones:2022upj}. For completeness, we provide in Appendix \ref{sec:appendix_B} a summary of the relevant equations of the EOS for point-like particles.
The mass of the quark quasiparticle of flavor $i$ is given by:
\begin{eqnarray}
M_i = m_i  +  \frac{C}{n_B^{a/3}} ,    \qquad  (i = u, d, s),
\label{eq:mass_formula_model2}
\end{eqnarray}
where $C$ and $a$ are positive flavor independent free parameters.

In order to calculate the energy density, we start from the expression for point-like particles given in Appendix \ref{sec:appendix_B}: 
\begin{equation}
\epsilon_\mathrm{pl} =  \sum_i g  M_i^4 \chi(x_i) ,
\end{equation}
where the variable $x_i$ is defined in Eq. \eqref{eq:appendix_B4} and the function $\chi$ is given in Eq. \eqref{eq:appendix_chi}.

Excluded volume effects are incorporated by writing the previous expression in terms of the set of variables $\{ n_j/q \}$ and including the correction factor $q$, as indicated in Eq. \eqref{eq:summary_epsilon_3fl}. To simplify the notation, we will use from now on the auxiliary variable 
\begin{equation}
\tilde{n}_i = \frac{n_i}{q(n_B)}.  
\end{equation}
After this procedure, the energy density reads:
\begin{equation}
\epsilon  =  \sum_i q(n_B) g \tilde{M}_i^4 \chi(\tilde{x}_i) ,
\label{eq:epsilon_QMDDM_3fl}
\end{equation}
where $g=6$ is the quark degeneracy and
\begin{eqnarray}
\tilde{M}_{i}  &\equiv &  M_{i}(\tilde{n}_B) =  m_{i}+\frac{C}{\tilde{n}_{B}^{a / 3}},  \\
\tilde{x}_i &\equiv &  x(\tilde{n}_i)  = \frac{1}{M_i(\tilde{n}_{B})} \left( \frac{6 \pi^2 \tilde{n}_i}{g} \right)^{1/3} , \\ 
\tilde{n}_B & \equiv &  N_B /\tilde{V}  =  \tfrac{1}{3} ( \tilde{n}_u  +  \tilde{n}_d  + \tilde{n}_s ) .
\end{eqnarray}
Adding the contribution of electrons to  Eq. \eqref{eq:epsilon_QMDDM_3fl} we obtain the complete energy density with excluded volume corrections:
\begin{equation}
\epsilon = \sum_{i=u, d, s}  q(n_B) g  \tilde{M}_i^4 \chi(\tilde{x}_i)  + \epsilon_e ,
\label{eq:final_epsilon_3fl}
\end{equation}
where electrons are described as point-like particles, i.e. $\epsilon_{e} = g_e  m_e^4  \chi(x_e)$,  $g_{e}=2$, $x_{e}=m_{e}^{-1} (6 \pi^{2} n_{e} / g_{e} )^{1 / 3}$, being $m_{e}$ the electron's mass.

To determine the pressure, we start from the expression for point-like particles given in Appendix \ref{sec:appendix_B}:
\begin{equation}
p_\mathrm{pl}  = \sum_i \left[ g  M_i^4 \phi(x_i) - B_i \right],
\label{eq:p_previous_paper_3fl}
\end{equation}
where $\phi$ is defined in Eq. \eqref{eq:appendix_phi}  and the ``bag constant" $B_i$ is given by Eq. \eqref{eq:bag_summary}.

To take into account excluded volume effects, we use Eq. \eqref{eq:summary_p_3fl}, i.e. we rewrite Eq. \eqref{eq:p_previous_paper_3fl} in terms of $\tilde{n}_i$ and add the correction factor $\delta$. The complete expression for the pressure is obtained adding the contribution of electrons. The result is: 
\begin{equation}
p = \delta(n_B)    \sum_{i=u, d, s}  \left[  g  \tilde{M}_i^4 \phi(\tilde{x}_i) - \tilde{B}_i \right]    + p_e ,
\label{eq:final_pressure_3fl}
\end{equation}
where $p_{e} = g_e  m_e^4  \phi(x_e)$ and   $\tilde{B}_i \equiv B(\tilde{n}_i)$.

Finally,  for determining the chemical potential we start from the expression for point-like particles given in Eq. \eqref{eq:mu_summary}:
\begin{equation}
\mu_{\mathrm{pl},i} = M_i \sqrt{x_i^2 + 1} - \frac{1}{3 n_B}  \sum_j B_j .
\label{eq:mu_previous_paper_3fl}
\end{equation}
To take into account excluded volume effects, we use Eq. \eqref{eq:summary_mu_3fl}, i.e. we rewrite Eqs. \eqref{eq:mu_previous_paper_3fl} and \eqref{eq:p_previous_paper_3fl} in terms of $\tilde{n}_i$ and include the correction factor $\lambda$. The result is:
\begin{equation}
\begin{aligned} 
\mu_i   = &  \lambda(n_B)    \sum_{i=u, d, s}  \left[  g  \tilde{M}_i^4 \phi(\tilde{x}_i) - \tilde{B}_i \right]   \\
& +  \tilde{M}_{i} \sqrt{\tilde{x}_{i}^{2}+1}-   \frac{1}{3 \tilde{n}_B}  \sum_{i=u, d, s}   \tilde{B}_i  .      \label{eq:chemical_3flavor}
\end{aligned}
\end{equation}

\section{3-flavor QMDDM with excluded volume effects: Numerical results}
\label{sec:3fl_numerical}

\subsection{The EOS}

Now, we adopt an ansatz for $b$ consistent with the asymptotic freedom behavior of QCD, i.e., 
we adopt a  formula that allows $b$ to approach zero at asymptotically large densities.  Based on the prescription of Eq. \eqref{eq:anzatz_decaying_b} with $\ell=1$, the excluded volume will be expressed as:
\begin{equation}
b = \frac{\kappa}{n_B} .
\label{eq:ansatz_figures}
\end{equation}
being $\kappa$ a positive constant. 

Using Eqs. \eqref{eq:summary_q_3fl}, \eqref{eq:summary_delta_3fl} and \eqref{eq:summary_lambda_3fl},  the functions $q$, $\delta$ and $\lambda$ are:
\begin{eqnarray}
q(n_B) &=&   1 - \kappa, \\
\delta(n_B) &=&  1 - \kappa, \\
\lambda(n_B) &= & 0 .
\end{eqnarray}

Replacing these parameters in Eqs. \eqref{eq:final_epsilon_3fl}, \eqref{eq:final_pressure_3fl} and \eqref{eq:chemical_3flavor} we obtain:
\begin{eqnarray}
\epsilon &=& \sum_{i=u, d, s}  (1 - \kappa) g  \tilde{M}_i^4 \chi(\tilde{x}_i)  + \epsilon_e ,   \label{eq:epsilon_figures} \\
p &=& \sum_{i=u, d, s}  (1 - \kappa)  (  g  \tilde{M}_i^4 \phi(\tilde{x}_i) - \tilde{B}_i)     + p_e ,  \label{eq:pressure_figures} \\
\mu_i   &=&  \tilde{M}_{i} \sqrt{\tilde{x}_{i}^{2}+1}-    \frac{1}{3 \tilde{n}_B}    \sum_{j}     \tilde{B}_{j}    .  
\label{eq:chemical_figures} 
\end{eqnarray}

\begin{figure}[b]
\centering 
\includegraphics[width=0.97\columnwidth]{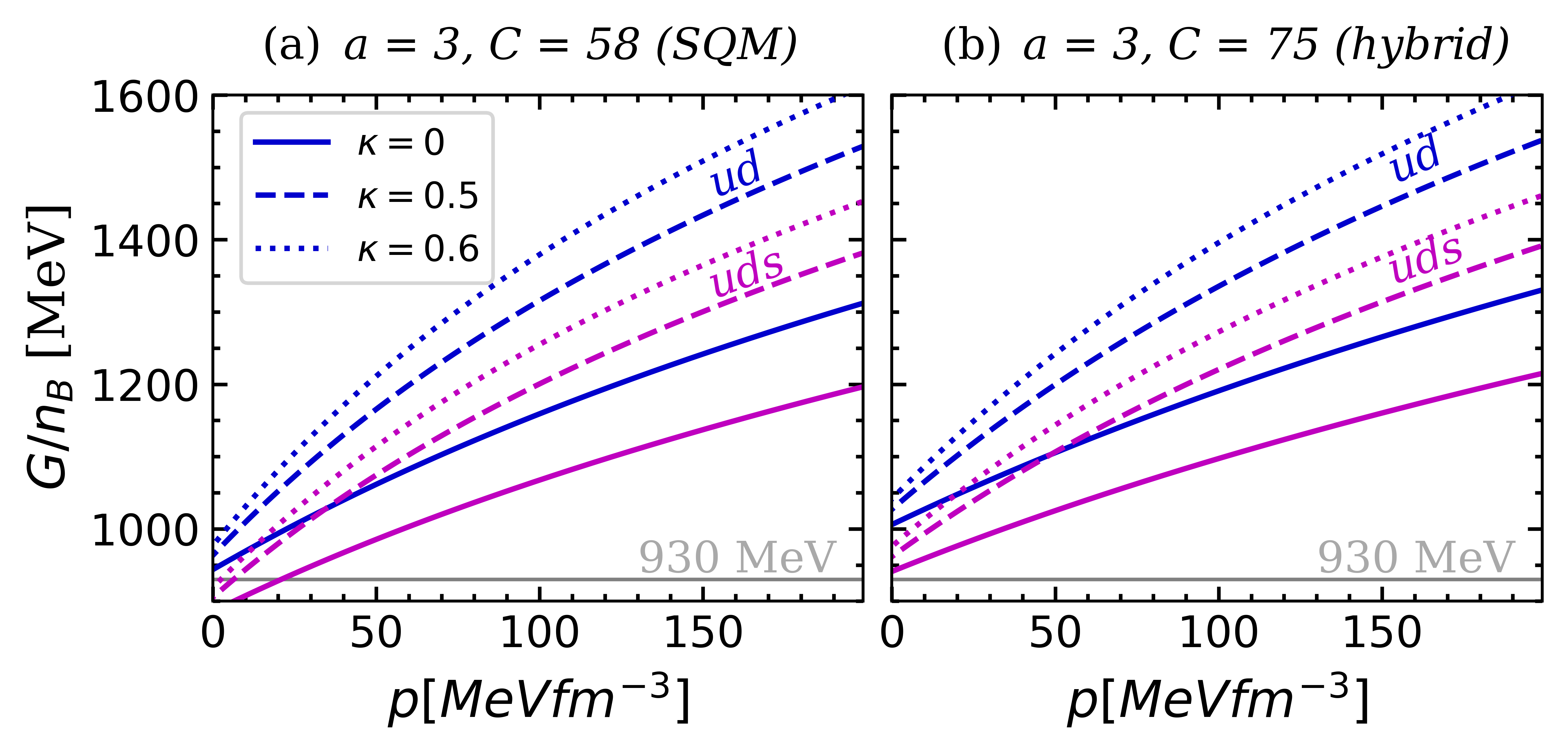}
\caption{Gibbs free energy per baryon of 2-flavor and 3-flavor quark matter in bulk for different choices of the EOS' parameters $a$, $C$ and $\kappa$. In panel (a), results are shown for a parameter choice leading to strange matter, while in panel (b) they are for a parameter choice yielding hybrid matter (see the main text for further details).}
\label{fig:3fl_gibbs}
\end{figure}

In the context of cold dense quark matter found in typical neutron star conditions, the preceding expressions need to be supplemented with the requirements of charge neutrality and chemical equilibrium under weak interactions. Assuming that neutrinos leave freely the system $\left(\mu_{\nu_{e}}=0\right)$,  the chemical equilibrium conditions read:
\begin{eqnarray}
\mu_{d} &=& \mu_{u}+\mu_{e}, \\
\mu_{s} &=& \mu_{d},    
\end{eqnarray}
while charge neutrality  is given by:
\begin{equation}
\tfrac{2}{3} n_{u}-\tfrac{1}{3} n_{d}-\tfrac{1}{3} n_{s}-n_{e}=0 .    
\end{equation}

\begin{figure}[tb]
\centering 
\includegraphics[width=0.97\columnwidth]{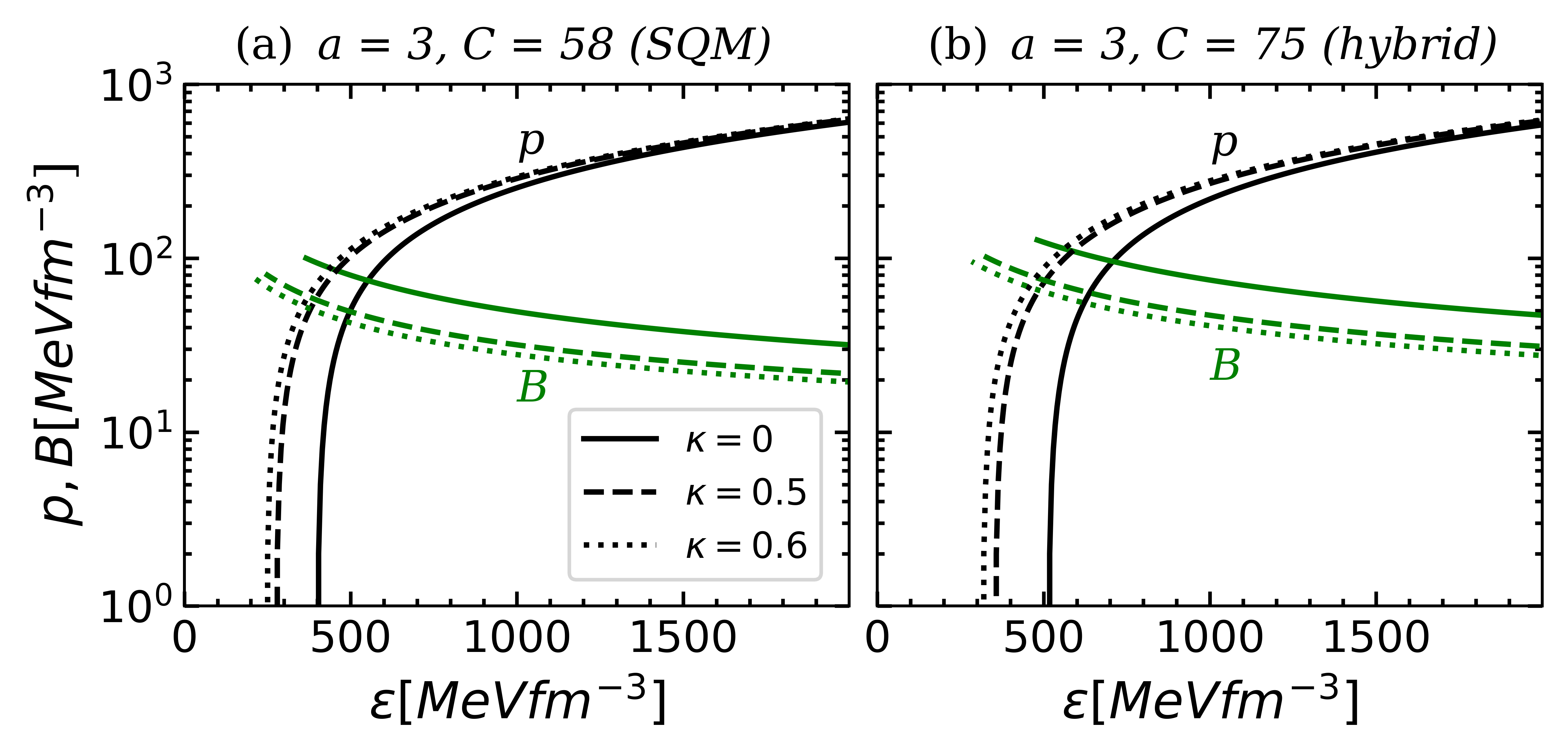}
\caption{Total pressure $p$ and  bag constant $B$ as a function of the energy density for the same parameter choices of  Fig. \ref{fig:3fl_gibbs}.}
\label{fig:3fl_pressure}
\end{figure}

\begin{figure}[tb]
\centering 
\includegraphics[width=0.97\columnwidth]{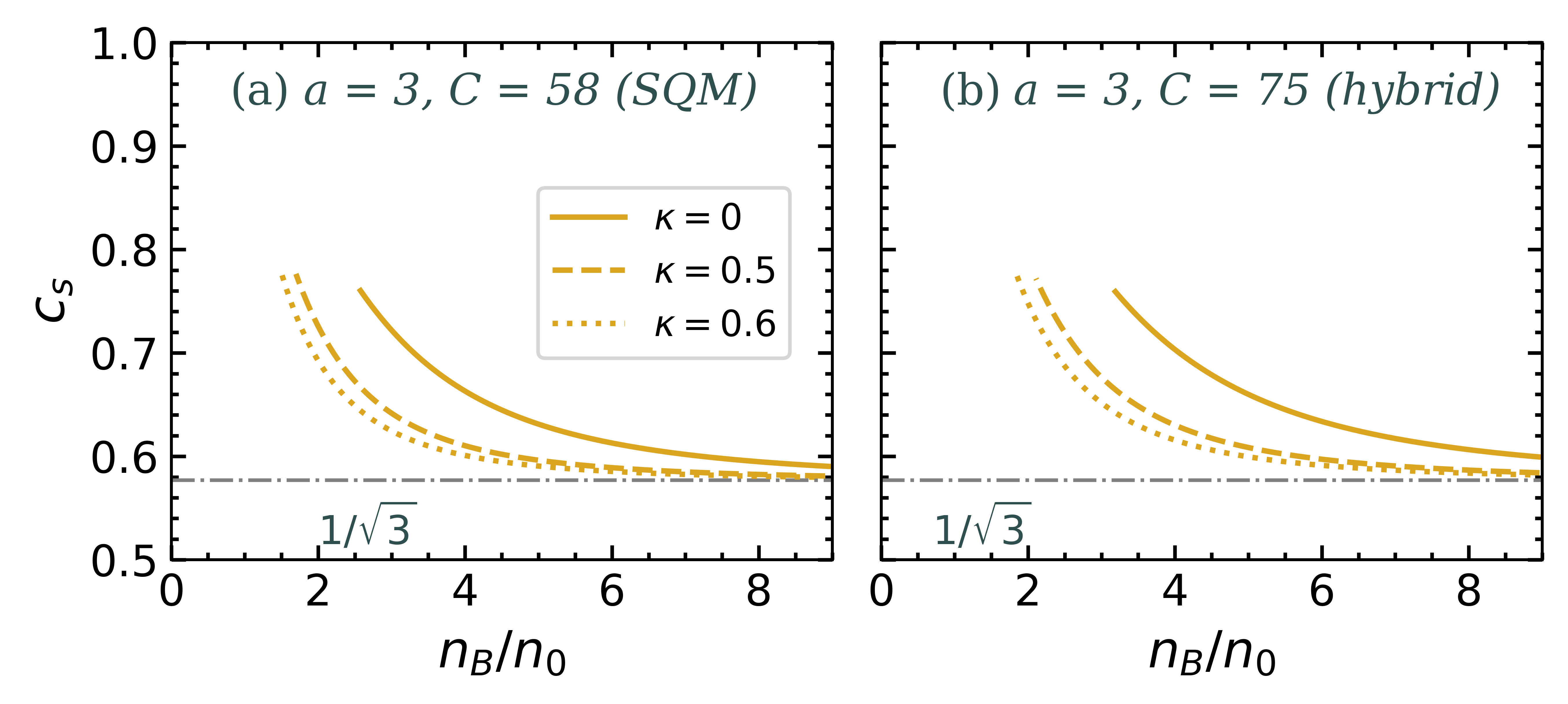}
\caption{Speed of sound  as a function of the baryon number density for the same parameter choices of  Fig. \ref{fig:3fl_gibbs}.}
\label{fig:3fl_sound}
\end{figure}

In Figs. \ref{fig:3fl_gibbs}, \ref{fig:3fl_pressure}, and \ref{fig:3fl_sound}, we present results for the EOS calculated with two different sets of parameters $a$ and $C$, along with three values of $\kappa$ (0, 0.5, 0.6).

In Fig. \ref{fig:3fl_gibbs}, we depict the Gibbs free energy per baryon, $G / n_{B} = (\epsilon+p) / n_{B}$, against pressure. Depending on the selected parameters $a$, $C$, and $\kappa$, the value of $G / n_{B}$ at $p=0$ can either exceed or remain below the energy per nucleon of the most tightly bound atomic nucleus, $^{62} \mathrm{Ni}$, approximately 930 MeV. Consequently, two distinct scenarios are presented in panels (a) and (b) of Fig. \ref{fig:3fl_gibbs}.
For the parameter set yielding $G / n_{B} < 930 \mathrm{MeV}$ at vanishing pressure and temperature, as illustrated in panel (a), we are in the domain of self-bound quark matter, i.e.  bulk quark matter in  vacuum remains stable and doesn't transition into hadronic matter. When this self-bound matter encompasses three flavors at $p=T=0$, it is designated as strange quark matter (SQM). Under these conditions, Nature would allow the existence of compact stars completely constituted of quark matter, known as self-bound quark stars\footnote{Note that, as shown in the analysis of Fig. 4 from Ref. \cite{Lugones:2022upj}, it is possible that two-flavor matter may be more stable than three-flavor matter for a specific choice of parameters. However, this specific case will not be analyzed in this study.}.
Conversely, panel (b) shows the situation where $G / n_{B} > 930 \mathrm{MeV}$. This is referred to as hybrid matter due to its transition from a hadronic state at lower pressures to a deconfined state at higher pressures. In such scenarios, stars with quark matter manifest as hybrid stars with a quark core surrounded by hadronic matter.
For the parameter selection illustrated in Fig. \ref{fig:3fl_gibbs}(b), the $uds$ curves consistently appear below their $ud$ counterparts, indicating that quark matter encompasses all three flavors. 
The Gibbs free energy curves exhibit significant sensitivity to variations in the parameter $\kappa$. Nonetheless, we always observe SQM in panel (a) and hybrid matter in panel (b) for all our choices of $\kappa$.

In Fig. \ref{fig:3fl_pressure}  we show the total pressure $p$ and the bag constant $B$ as a function of the energy density for the same parameter choices of  Fig. \ref{fig:3fl_gibbs}. In all cases the pressure becomes negative at finite energy density due to the effect of $B$. The bag constant depends on density, always being a decreasing function of $\epsilon$. At asymptotically large densities $B$ tends to zero and the system behaves as a free Fermi gas of electrons and quarks with $M_{i}=m_{i}$. 
The EOS is notably sensitive to changes in the parameter $\kappa$, invariably leading to an increase in the stiffness of the EOS as $\kappa$ increases. As $\kappa$ rises, the bag constant decreases, which in turn shifts the energy density at which the matter pressure becomes zero to lower energy density values. Since the excluded volume is inversely dependent on the baryon number density, the excluded volume effect vanishes at high densities, causing the curves with different $\kappa$ values to converge  with each other.

The speed of sound $c_{s}$ is shown in Fig. \ref{fig:3fl_sound} for the same parameter choices of previous figures. In all cases $c_{s}$ is a decreasing function of the baryon number density and tends asymptotically to the conformal limit $c_{s}=1 / \sqrt{3}$.  Since the pressure tends to zero at a finite density, the curves are truncated at the point where $p = 0$. As the value of the parameter $\kappa$ increases, the speed of sound decreases for a given density. This difference vanishes at asymptotically high densities, at which point all curves converge with each other.

\subsection{Astrophysical applications}

Finally, we analyze stellar configurations based on the two main parameter sets discussed in the previous section, representing strange quark matter and hybrid matter. This paper does not aim for a comprehensive investigation of stellar structure; a more detailed exploration is planned for a future publication. Our primary goal here is to show that the presented model can produce stellar configurations, both strange and hybrid, consistent with current astrophysical constraints.

\begin{figure}[tb]
\centering 
\includegraphics[width=0.9\columnwidth]{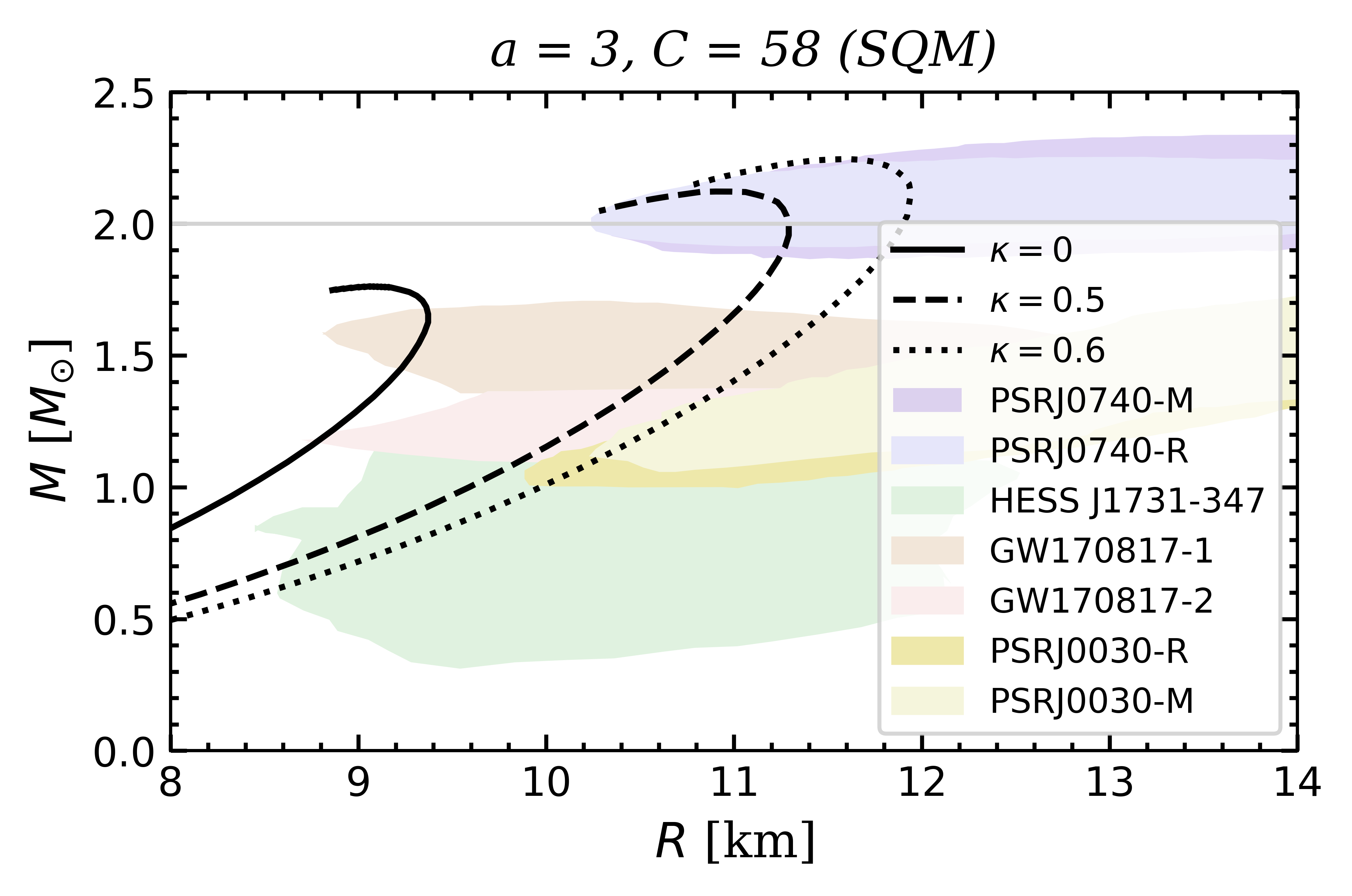}
\caption{Mass-radius relationship for strange quark stars with parameters $a = 3$ and $C = 58$. Curves represent different values of the parameter $\kappa$. Colored bands correspond to  the  $95\%$ confidence intervals for the mass and radius of the millisecond-pulsars PSR J0030+0451 \cite{Riley:2019yda,Miller:2019cac} and PSR J0740+6620 \cite{Riley:2021pdl,Miller:2021qha} measured recently by the Neutron Star Interior Composition Explorer (NICER), and the $90\%$ confidence intervals for the merging event GW170817 \cite{Abbott:2018exr} (LIGO/Virgo).}
\label{fig:strange_stars}
\end{figure}

Let us consider first strange star configurations shown in Fig. \ref{fig:strange_stars}. For point-like quarks ($\kappa = 0$), the maximum mass does not reach the required  $2 M_{\odot}$ constraint and the mass-radius curve does not fulfill the astrophysical constraints set by NICER \cite{Riley:2019yda,Miller:2019cac,Riley:2021pdl,Miller:2021qha} observations. Upon accounting for the excluded volume of quarks, the EOS stiffens as described in the previous section. Consequently, the maximum mass rises, and the stellar radii increase for a given mass. The maximum mass increases significantly, approaching close to $2.3 M_{\odot}$ for $\kappa = 0.6$. 


\begin{figure}[tb]
\centering 
\includegraphics[width=0.9\columnwidth]{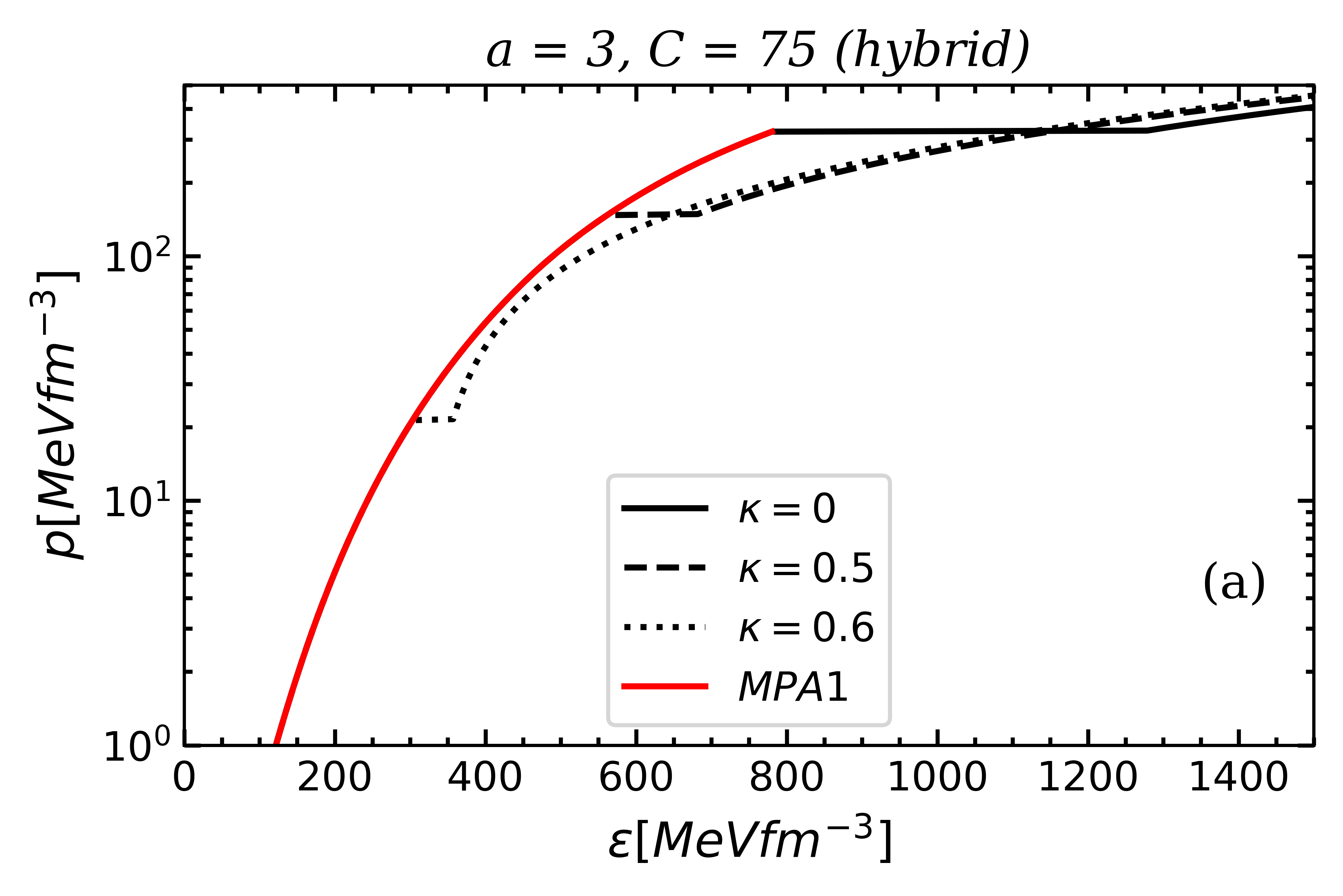}
\includegraphics[width=0.9\columnwidth]{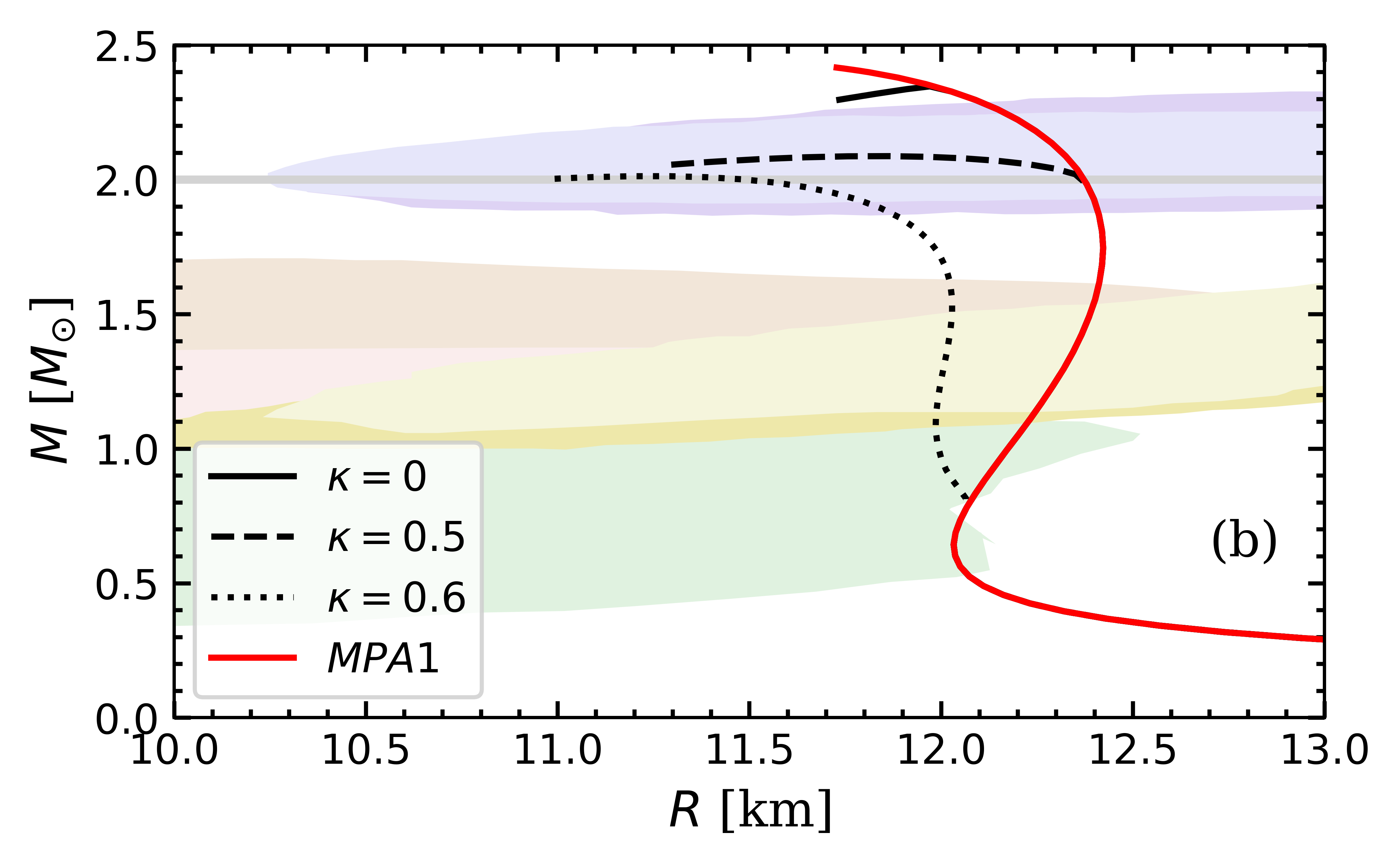}
\caption{Panel (a) illustrates the EOS for hybrid matter with parameters $a = 3$ and $C = 75$. The red curve represents the hadron matter EOS according to the MPA1 model. The black curves depict the quark matter EOS for point-like quasiparticles ($\kappa = 0$), and with excluded volume effects with $\kappa = 0.5$ and $\kappa = 0.6$.  Panel (b) shows the corresponding mass-radius relationship for these EOS. The colored bands in the lower diagram denote the observational constraints introduced in Fig. \ref{fig:strange_stars}.}
\label{fig:hybrid_stars}
\end{figure}

In Fig. \ref{fig:hybrid_stars}  we analyze  the properties of the hybrid matter EOS and its implications on the mass-radius relationship of compact objects.
The hybrid EOS shown in the upper panel was constructed using the relativistic Brueckner-Hartree-Fock hadronic  EOS  known as MPA1 \cite{Muther:1987xaa} (red curve) together with quark  matter characterized by parameters $a = 3$ and $C = 75$, with different levels of volume exclusion (black curves).  The plateaus represent sharp first-order phase transitions between hadrons and quarks. Increasing $\kappa$ diminishes the transition pressure as well as the energy density jump between both phases. The quark EOS becomes stiffer as $\kappa$ increases.  
In panel (b) we show the mass-radius relationship, resulting from the aforementioned EOS. 
It should be noted that for quark matter composed of point-like particles ($\kappa =0$), all dynamically stable configurations (i.e. to the right of the maximum mass point) are hadronic.  Hybrid stars appear only beyond the maximum mass object and are dynamically unstable for $\kappa =0$. 
As the transition pressure decreases with an increasing value of $\kappa$, it becomes possible for dynamically stable hybrid stellar configurations to emerge, satisfying modern astrophysical constraints.
Notice that the maximum mass decreases as  $\kappa$ increases. Specifically, for the value $\kappa = 0.5$, the maximum mass is around $2.1 M_{\odot}$, and for $\kappa = 0.6$ it is around $2 M_{\odot}$. It is important to emphasize that finding parameterizations of the EOS that yield hybrid stars with $M_\mathrm{max} > 2 M_{\odot}$ is challenging when considering point-like particles. In this regard, the inclusion of excluded volume effects is essential for the emergence of dynamically stable models of hybrid stars consistent with astrophysical requirements.


The main conclusion from the preceding analysis is that the QMDDM with excluded volume effects is not only  viable from an astrophysical standpoint, but that excluded volume effects play a crucial role to make the model consistent with astrophysical constraints.   A more in-depth examination is reserved for a future study.

\section{Summary and conclusions}

In our previous study \cite{Lugones:2022upj}, we revisited the QMDDM model, adopting the canonical ensemble framework in place of the usual grand canonical ensemble. This methodological change was essential for resolving the thermodynamic inconsistencies that had persisted in the model for decades.
The QMDDM is based on the idea that some relevant aspects of the strong interaction between quarks, particularly in the high-density, low-temperature regime, can be effectively modeled by treating the system as a gas of quasiparticles, wherein the effective masses of these quarks depend on the local density. As a natural consequence of this density-dependent quark mass, a ``bag'' term naturally arises in the pressure, resulting in quark confinement. Clearly, not all qualitative features of strong interactions can be faithfully replicated through the simple introduction of density dependent masses. Improved versions of the model are necessary to better capture some phenomenological aspects of non-perturbative QCD. In this present study, we focused our efforts in that direction, incorporating the influence of repulsive interactions. We approached this by departing from the assumption of point-like particles and instead considering particles as entities endowed with an excluded volume in their vicinity. The present approach aims to bring a more realistic representation of the quark matter EOS of dense matter at vanishing temperature.

To account for the influence of excluded volume, we introduced the ``available volume", $\tilde{V}=V-b(n) N$, in the Helmholtz free energy $F_\mathrm{pl}(V, N)$ for point-like particles, where $V$ is the system's volume. The parameter $b$, which can be a function of the particle's number density $n$, represents the volume that each particle would exclude if it were considered a rigid sphere.
Once $V$ is replaced with $\tilde{V}$, the resulting function $F_\mathrm{pl}(\tilde{V}, N)$ becomes the starting point for describing the system with excluded volume effects. 

In Secs. \ref{sec:generic_1fl} and \ref{sec:1fl_QMDDM} we focused in a simple one-flavor model and showed that it is  possible to incorporate  excluded volume corrections into the existing expressions of point-like  particles in a straightforward manner, achieving a quite practical approach. The procedure involves substituting the particle density $n$ that appears in the expressions for point-like particles with a density $n/q$ where the correction factor $q$ is defined in Eq. \eqref{eq:summary_q_1fl} and takes into account the available volume in the system. Additionally, multiplicative correction factors arise in the energy density (Eq. \eqref{eq:summary_epsilon_1fl}) pressure (Eq. \eqref{eq:summary_p_1fl}), and the chemical potential (Eq. \eqref{eq:summary_mu_1fl}).
Using the aforementioned equations, we analyzed two different ansatzes for the excluded volume. Initially, we considered a scenario where particles always exclude the same volume regardless of the matter density. This assumption, as depicted Fig. \ref{fig:constantv0_EOS_1}(d) is not satisfactory as the speed of sound becomes acausal with increasing density. A more adequate prescription must account for the asymptotic freedom of particles, requiring them to behave as free point-like particles as density increases. To incorporate this behavior, we adopted an ansatz for the excluded volume that is inversely proportional to density, causing it to decrease and tend toward zero as density approaches infinity. This assumption cures the unphysical behavior of the speed of sound, causing it to asymptotically approach the conformal limit $c_{s}=1 / \sqrt{3}$, as illustrated in Fig. \ref{fig:variable_b_EOS}(d). Moreover, the EOS becomes significantly stiffer, as shown in  Fig. \ref{fig:variable_b_EOS}(c).

In Secs. \ref{sec:generic_3fl} and \ref{sec:3fl_QMDDM},  we extended the EOS to a three-flavor system. In this context, we assumed that both the mass and the excluded volume depend on the baryon number density. Similar to the single-flavor case, we can directly incorporate the excluded volume corrections into the existing EOS for point-like particles, as summarized in Sec. \ref{sec:summary_generic_3fl}.  By applying these formulas to the model outlined in  Ref.~\cite{Lugones:2022upj}, we derived explicit expressions for the energy density (Eq.~\eqref{eq:final_epsilon_3fl}), pressure (Eq.~\eqref{eq:final_pressure_3fl}), and chemical potential (Eq.~\eqref{eq:chemical_3flavor}) of a system composed by quarks $u$, $d$, $s$, and electrons.

In Sec. \ref{sec:3fl_numerical} we adopted an ansatz for the excluded volume that mimics the asymptotic freedom behavior of QCD. Similarly to the single-species case we assumed that the excluded volume per baryon is inversely proportional to the baryon number density (Eq.~\eqref{eq:ansatz_figures}). For this specific choice, the EOS takes the form given in Eqs.~\eqref{eq:epsilon_figures}, \eqref{eq:pressure_figures}, and \eqref{eq:chemical_figures}.
Given our focus on astrophysical applications, our numerical computations of the EOS are centered on an electrically neutral system in chemical equilibrium under weak interactions. To represent the two most relevant astrophysical scenarios, we adopt  sets of parameters $a$ and $C$, to characterize self-bound matter and hybrid matter.  For the parameter $\kappa$, we adopt values that correspond to different excluded volume sizes.

In both the self-bound and hybrid cases, based on the chosen parameter set, the $uds$ composition consistently exhibits the lower $G / n_{B}$, being energetically preferred. It's worth noting that the Gibbs free energy curves display a remarkable sensitivity to changes in the parameter $\kappa$ (see Fig. \ref{fig:3fl_gibbs}).
The EOS is substantially affected by changes in the parameter $\kappa$,  leading to an increase in the stiffness as $\kappa$ increases. At high densities the excluded volume effect vanishes and the curves with different $\kappa$ values converge with each other (see Fig. \ref{fig:3fl_pressure}). The speed of sound (see Fig. \ref{fig:3fl_sound}) exhibits a decreasing trend with increasing baryon number density, ultimately approaching the conformal limit asymptotically. At a specific density, the excluded volume effect tends to diminish the speed of sound. However, this distinction becomes negligible at very high densities, where all curves converge with each other.

Finally, we studied stellar configurations for the two parameter sets representing self-bound and hybrid matter. In the first case, shown in Fig. \ref{fig:strange_stars}, we found that stellar configurations are compatible with all current astrophysical constraints only for a significant amount of volume exclusion. Since the EOS becomes remarkably stiffer as excluded volume increases, the maximum mass of self-bound stars significantly rises for large $\kappa$ approaching close to $2.3 M_{\odot}$ for $\kappa = 0.6$.  Recently, evidence has been found that suggests that the extremely tiny and light compact object, HESS J1731-347, may in fact be a self-bound star with a mass around $0.77 M_{\odot}$ and a radius of $\sim 10 \, \mathrm{km}$ \cite{HESS2022}. As shown in Fig. \ref{fig:strange_stars} the mass-radius curves of self-bound objects naturally cross the credibility contours of HESS J1731-347 if a substantial amount of excluded volume is incorporated in the EOS. If confirmed, this object could be a strange star candidate.  We also explored hybrid star configurations using a representative hadronic EOS and various parametrizations of the QMDDM. From our calculations we learn that  the constraint of $2 M_{\odot}$ is fulfilled only for very stiff hadronic EOS together with a QMDDM EOS incorporating a substantial degree of volume exclusion. In Fig. \ref{fig:hybrid_stars} it can be seen that the hybrid curves composed by the MPA1 hadronic EOS and the QMDDM with $\kappa = 0.5$ and $0.6$ are in agreement with all the modern multimessenger constraints.   

In conclusion, we have extended the QMDDM model, which was previously modified in an earlier work to ensure thermodynamic consistency, to incorporate the effects of excluded volume. These effects phenomenologically represent repulsive interactions between quasiparticles. The incorporation of these effects leads to stiffer EOS increasing the masses of compact objects, aligning them more closely with recent astrophysical observations. A comprehensive analysis of stellar structure using this model is planned as the focus of our future research.

\section{Acknowledgements}

G.L. acknowledges the financial support from the Brazilian agencies CNPq (grant 316844/2021-7) and FAPESP (grant 2022/02341-9). A. G. G. would like to acknowledge the financial support from CONICET under Grant No. PIP 22-24 11220210100150CO,  ANPCyT (Argentina) under Grant PICT20-01847, and the National University of La Plata (Argentina), Project No. X824.

\appendix

\section{Summary of the 1-flavor QMDDM EOS for point-like particles}
\label{sec:appendix_A}

For the sake of completeness, we summarize below all the expressions derived in Ref. \cite{Lugones:2022upj} for the QMDDM in the one-flavor case. As discussed in the main text, these expressions describe a system of \textit{point-like} particles. The Helmholtz free energy is given by:
\begin{equation}
F =  g V M^{4} \chi(x) , 
\end{equation}
where
\begin{eqnarray}
M(n) & = &   m + \frac{C}{n^{a/3}}  , \\
x(n) & = & \frac{1}{M}\left(\frac{6 \pi^{2} n}{g}\right)^{1 / 3} , \label{eq:appendix_A3} \\
\chi(x)  & =  &  \frac{ x \sqrt{x^2 + 1} (2 x^2 + 1) - \arcsinh(x)  }{16 \pi^2}, \label{eq:appendix_chi}
\end{eqnarray}
being $g$ the particle's degeneracy, $M$ the density dependent mass, and $x$ the dimensionless Fermi momentum.

The energy density at $T=0$ is given by $\epsilon=F / V$ and reads:
\begin{align}
\epsilon(n)=g M^{4} \chi(x) .
\label{eq:appendix_epsilon}
\end{align}

The pressure is given by
\begin{equation}
p(n) = n^{2} \frac{\partial(\epsilon / n)}{\partial n}  ,
\label{eq:appendix_p1}
\end{equation}
which results in:
\begin{equation}
p(n) =  p^{\mathrm{FG}}(n) - B(n) ,
\label{eq:appendix_p2}
\end{equation}
being:
\begin{eqnarray} 
p^{\mathrm{FG}}(n) & = & g M^{4} \phi(x) , \\ 
B(n) & = &  -g M^{3} n \frac{\partial M }{\partial n } \beta(x)>0 ,  \label{eq:appendix_bag} \\
\frac{\partial M }{\partial n } & =  & -\frac{C}{3} \frac{a}{n^{a / 3+1}}  ,   \\
\phi(x) &  = &  \frac{  x \sqrt{x^{2}+1}(2 x^{2} -3) +3 \operatorname{arcsinh}(x)  }{48 \pi^{2}} , \quad   \label{eq:appendix_phi} \\
\beta(x) & = &  \frac{1}{4 \pi^{2}}\left[x \sqrt{x^{2}+1} -\operatorname{arcsinh}(x)\right] \label{eq:appendix_beta} .
\end{eqnarray}
The quantity $B$ serves as a bag constant, inducing negative pressure at sufficiently low densities, mimicking the effects of quark confinement.

Finally, the chemical potential is:
\begin{align}
\mu(n) =  \frac{\partial \epsilon(n) }{\partial n} =  g \frac{\partial  [M^4(n) \chi(x)] }{\partial n} .
\label{eq:appendix_mu1}
\end{align}
As shown in Eq. (40) of Ref. \cite{Lugones:2022upj} the latter derivative results in:
\begin{align}
\mu(n) = \mu^{\mathrm{FG}}(n)-\frac{B(n)}{n},
\label{eq:appendix_mu2}
\end{align}
with
\begin{align}
\mu^{\mathrm{FG}}(n)=M \sqrt{x^{2}+1}. 
\label{eq:appendix_mu_FG}
\end{align}

\section{Summary of the 3-flavor QMDDM EOS for point-like particles with a flavor-blind mass formula}
\label{sec:appendix_B}

In this appendix, we present a synopsis of the QMDDM EOS for point-like quarks, as developed in Ref. \cite{Lugones:2022upj}. We concentrate on the parametrization of the QMDDM with a flavor-independent mass formula. Specifically,  we assume that the mass of the quark quasiparticle of flavor $i$ depends on the baryon number density $n_B$ as follows:
\begin{eqnarray}
M_i = m_i  +  \frac{C}{n_B^{a/3}} ,    \qquad  (i = u, d, s),
\end{eqnarray}
where $C$ and $a$ are free parameters, and $n_B = \tfrac{1}{3} (n_u + n_d + n_s)$.
The flavor dependence of $M_i$ comes only from the different values of the current masses $m_i$.

We will describe the system as a mixture of non-interacting quarks with effective masses $M_i$ 
and free electrons. The total Helmholtz free energy is simply the sum of the contribution of each species:
\begin{equation}
F(V,n_B) = \sum_{i=u, d, s, e}  F_i . 
\end{equation}
where, 
\begin{equation}
F_i  =
\begin{cases} g V M_i^4 \chi(x_i) & \quad (i = u, d, s) , \\ 
 g_e V m_e^4 \chi(x_e)  & \quad  (\text{electrons}) ,
\end{cases}
\label{eq:F_i_summary}
\end{equation}
with $g=6$ and $g_e =2$.  The function $\chi(x)$ is defined in Eq. \eqref{eq:appendix_chi} and: 
\begin{eqnarray}
x_i  = \frac{1}{M_i} \left( \frac{6 \pi^2 n_i}{g} \right)^{1/3} & \quad (i = u, d, s) ,    \label{eq:appendix_B4} \\ 
x_e  =  \frac{1}{m_e} \left( \frac{6 \pi^2 n_e}{g_e} \right)^{1/3} & \quad  (\text{electrons}) .   \label{eq:appendix_B5}
\end{eqnarray}
The energy density is $\epsilon = \sum_{i=u, d, s, e}  \epsilon_i$, where:
\begin{equation}
\epsilon_i  =
\begin{cases} g M_i^4 \chi(x_i) & \quad (i = u, d, s) , \\ 
 g_e m_e^4 \chi(x_e)  & \quad (\text{electrons}) .
\end{cases}
\label{eq:epsilon_i_summary}
\end{equation}
The total  pressure is $p = \sum_{i=u, d, s, e}  p_i$, being
\begin{equation}
p_i  =
\begin{cases} g M_i^4 \phi(x_i) - B_i  & \quad (i = u, d, s) , \\ 
  g_e m_e^4 \phi(x_e)  & \quad (\text{electrons}) ,
\end{cases}
\label{eq:p_i_summary}
\end{equation}
where $\phi(x)$ is defined in Eq. \eqref{eq:appendix_phi} and  the ``bag constant'' $B_i$ is given by: 
\begin{equation}
\begin{aligned}
- B_i & =   g n_i  M_i^3  \frac{\partial M_i}{\partial n_i} \beta(x_i) \\
      & =  g n_B  M_i^3  \frac{\partial M_i}{\partial n_B} \beta(x_i),
\label{eq:bag_summary}
\end{aligned}
\end{equation}
with $\beta(x)$ defined in Eq. \eqref{eq:appendix_beta}. In our previous paper, only the first line of Eq. \eqref{eq:bag_summary} was presented. The second line is simpler to compute in the flavor-blind model, and it is straightforward to demonstrate the equivalence of both lines.

Finally, the chemical potential of $u$, $d$ and $s$ quasiparticles is: 
\begin{equation}
\mu_i  =  M_i \sqrt{x_i^{2}+1} -   \frac{1}{3 n_B} \sum_{j}  B_j  ,
\label{eq:mu_summary}
\end{equation}
while for electrons it reads:
\begin{eqnarray}
\mu_e = m_e \sqrt{x_e^{2}+1} . 
\label{eq:mu_summary_2}
\end{eqnarray}
It is important to note that the second term of  Eq. \eqref{eq:mu_summary} differs from the expression presented in Eq. (88) of Ref. \cite{Lugones:2022upj} due to an error in that earlier work.

\bibliography{references}

\begin{thebibliography}{29}%
\makeatletter
\providecommand \@ifxundefined [1]{%
 \@ifx{#1\undefined}
}%
\providecommand \@ifnum [1]{%
 \ifnum #1\expandafter \@firstoftwo
 \else \expandafter \@secondoftwo
 \fi
}%
\providecommand \@ifx [1]{%
 \ifx #1\expandafter \@firstoftwo
 \else \expandafter \@secondoftwo
 \fi
}%
\providecommand \natexlab [1]{#1}%
\providecommand \enquote  [1]{``#1''}%
\providecommand \bibnamefont  [1]{#1}%
\providecommand \bibfnamefont [1]{#1}%
\providecommand \citenamefont [1]{#1}%
\providecommand \href@noop [0]{\@secondoftwo}%
\providecommand \href [0]{\begingroup \@sanitize@url \@href}%
\providecommand \@href[1]{\@@startlink{#1}\@@href}%
\providecommand \@@href[1]{\endgroup#1\@@endlink}%
\providecommand \@sanitize@url [0]{\catcode `\\12\catcode `\$12\catcode
  `\&12\catcode `\#12\catcode `\^12\catcode `\_12\catcode `\%12\relax}%
\providecommand \@@startlink[1]{}%
\providecommand \@@endlink[0]{}%
\providecommand \url  [0]{\begingroup\@sanitize@url \@url }%
\providecommand \@url [1]{\endgroup\@href {#1}{\urlprefix }}%
\providecommand \urlprefix  [0]{URL }%
\providecommand \Eprint [0]{\href }%
\providecommand \doibase [0]{https://doi.org/}%
\providecommand \selectlanguage [0]{\@gobble}%
\providecommand \bibinfo  [0]{\@secondoftwo}%
\providecommand \bibfield  [0]{\@secondoftwo}%
\providecommand \translation [1]{[#1]}%
\providecommand \BibitemOpen [0]{}%
\providecommand \bibitemStop [0]{}%
\providecommand \bibitemNoStop [0]{.\EOS\space}%
\providecommand \EOS [0]{\spacefactor3000\relax}%
\providecommand \BibitemShut  [1]{\csname bibitem#1\endcsname}%
\let\auto@bib@innerbib\@empty
\bibitem [{\citenamefont {Lugones}\ and\ \citenamefont
  {Grunfeld}(2023)}]{Lugones:2022upj}%
  \BibitemOpen
  \bibfield  {author} {\bibinfo {author} {\bibfnamefont {G.}~\bibnamefont
  {Lugones}}\ and\ \bibinfo {author} {\bibfnamefont {A.~G.}\ \bibnamefont
  {Grunfeld}},\ }\bibfield  {title} {\bibinfo {title} {{Cold dense quark matter
  with phenomenological medium effects: A self-consistent formulation of the
  quark-mass density-dependent model}},\ }\href
  {https://doi.org/10.1103/PhysRevD.107.043025} {\bibfield  {journal} {\bibinfo
   {journal} {Phys. Rev. D}\ }\textbf {\bibinfo {volume} {107}},\ \bibinfo
  {pages} {043025} (\bibinfo {year} {2023})},\ \Eprint
  {https://arxiv.org/abs/2209.03455} {arXiv:2209.03455 [nucl-th]} \BibitemShut
  {NoStop}%
\bibitem [{\citenamefont {Annala}\ \emph {et~al.}(2020)\citenamefont {Annala},
  \citenamefont {Gorda}, \citenamefont {Kurkela}, \citenamefont {N\"attil\"a},\
  and\ \citenamefont {Vuorinen}}]{Annala:2019puf}%
  \BibitemOpen
  \bibfield  {author} {\bibinfo {author} {\bibfnamefont {E.}~\bibnamefont
  {Annala}}, \bibinfo {author} {\bibfnamefont {T.}~\bibnamefont {Gorda}},
  \bibinfo {author} {\bibfnamefont {A.}~\bibnamefont {Kurkela}}, \bibinfo
  {author} {\bibfnamefont {J.}~\bibnamefont {N\"attil\"a}},\ and\ \bibinfo
  {author} {\bibfnamefont {A.}~\bibnamefont {Vuorinen}},\ }\bibfield  {title}
  {\bibinfo {title} {{Evidence for quark-matter cores in massive neutron
  stars}},\ }\href {https://doi.org/10.1038/s41567-020-0914-9} {\bibfield
  {journal} {\bibinfo  {journal} {Nature Phys.}\ }\textbf {\bibinfo {volume}
  {16}},\ \bibinfo {pages} {907} (\bibinfo {year} {2020})},\ \Eprint
  {https://arxiv.org/abs/1903.09121} {arXiv:1903.09121 [astro-ph.HE]}
  \BibitemShut {NoStop}%
\bibitem [{\citenamefont {Fowler}\ \emph {et~al.}(1981)\citenamefont {Fowler},
  \citenamefont {Raha},\ and\ \citenamefont {Weiner}}]{Fowler:1981rp}%
  \BibitemOpen
  \bibfield  {author} {\bibinfo {author} {\bibfnamefont {G.~N.}\ \bibnamefont
  {Fowler}}, \bibinfo {author} {\bibfnamefont {S.}~\bibnamefont {Raha}},\ and\
  \bibinfo {author} {\bibfnamefont {R.~M.}\ \bibnamefont {Weiner}},\ }\bibfield
   {title} {\bibinfo {title} {{Confinement and Phase Transitions}},\ }\href
  {https://doi.org/10.1007/BF01410668} {\bibfield  {journal} {\bibinfo
  {journal} {Z. Phys. C}\ }\textbf {\bibinfo {volume} {9}},\ \bibinfo {pages}
  {271} (\bibinfo {year} {1981})}\BibitemShut {NoStop}%
\bibitem [{\citenamefont {Chakrabarty}\ \emph {et~al.}(1989)\citenamefont
  {Chakrabarty}, \citenamefont {Raha},\ and\ \citenamefont
  {Sinha}}]{Chakrabarty:1989bq}%
  \BibitemOpen
  \bibfield  {author} {\bibinfo {author} {\bibfnamefont {S.}~\bibnamefont
  {Chakrabarty}}, \bibinfo {author} {\bibfnamefont {S.}~\bibnamefont {Raha}},\
  and\ \bibinfo {author} {\bibfnamefont {B.}~\bibnamefont {Sinha}},\ }\bibfield
   {title} {\bibinfo {title} {{Strange Quark Matter and the Mechanism of
  Confinement}},\ }\href {https://doi.org/10.1016/0370-2693(89)90166-4}
  {\bibfield  {journal} {\bibinfo  {journal} {Phys. Lett. B}\ }\textbf
  {\bibinfo {volume} {229}},\ \bibinfo {pages} {112} (\bibinfo {year}
  {1989})}\BibitemShut {NoStop}%
\bibitem [{\citenamefont {Chakrabarty}(1991)}]{Chakrabarty:1991ui}%
  \BibitemOpen
  \bibfield  {author} {\bibinfo {author} {\bibfnamefont {S.}~\bibnamefont
  {Chakrabarty}},\ }\bibfield  {title} {\bibinfo {title} {{Equation of state of
  strange quark matter and strange star}},\ }\href
  {https://doi.org/10.1103/PhysRevD.43.627} {\bibfield  {journal} {\bibinfo
  {journal} {Phys. Rev. D}\ }\textbf {\bibinfo {volume} {43}},\ \bibinfo
  {pages} {627} (\bibinfo {year} {1991})}\BibitemShut {NoStop}%
\bibitem [{\citenamefont {Chakrabarty}(1993)}]{Chakrabarty:1993db}%
  \BibitemOpen
  \bibfield  {author} {\bibinfo {author} {\bibfnamefont {S.}~\bibnamefont
  {Chakrabarty}},\ }\bibfield  {title} {\bibinfo {title} {{Stability of strange
  quark matter at T not = 0}},\ }\href
  {https://doi.org/10.1103/PhysRevD.48.1409} {\bibfield  {journal} {\bibinfo
  {journal} {Phys. Rev. D}\ }\textbf {\bibinfo {volume} {48}},\ \bibinfo
  {pages} {1409} (\bibinfo {year} {1993})}\BibitemShut {NoStop}%
\bibitem [{\citenamefont {Benvenuto}\ and\ \citenamefont
  {Lugones}(1995)}]{Benvenuto:1989kc}%
  \BibitemOpen
  \bibfield  {author} {\bibinfo {author} {\bibfnamefont {O.~G.}\ \bibnamefont
  {Benvenuto}}\ and\ \bibinfo {author} {\bibfnamefont {G.}~\bibnamefont
  {Lugones}},\ }\bibfield  {title} {\bibinfo {title} {{Strange matter equation
  of state in the quark mass density dependent model}},\ }\href
  {https://doi.org/10.1103/PhysRevD.51.1989} {\bibfield  {journal} {\bibinfo
  {journal} {Phys. Rev.}\ }\textbf {\bibinfo {volume} {D51}},\ \bibinfo {pages}
  {1989} (\bibinfo {year} {1995})}\BibitemShut {NoStop}%
\bibitem [{\citenamefont {Lugones}\ and\ \citenamefont
  {Benvenuto}(1995)}]{Lugones:1995vg}%
  \BibitemOpen
  \bibfield  {author} {\bibinfo {author} {\bibfnamefont {G.}~\bibnamefont
  {Lugones}}\ and\ \bibinfo {author} {\bibfnamefont {O.~G.}\ \bibnamefont
  {Benvenuto}},\ }\bibfield  {title} {\bibinfo {title} {{Strange matter
  equation of state and the combustion of nuclear matter into strange matter in
  the quark mass density dependent model at $T > 0$}},\ }\href
  {https://doi.org/10.1103/PhysRevD.52.1276} {\bibfield  {journal} {\bibinfo
  {journal} {Phys. Rev.}\ }\textbf {\bibinfo {volume} {D52}},\ \bibinfo {pages}
  {1276} (\bibinfo {year} {1995})}\BibitemShut {NoStop}%
\bibitem [{\citenamefont {Benvenuto}\ and\ \citenamefont
  {Lugones}(1998)}]{Benvenuto:1998tx}%
  \BibitemOpen
  \bibfield  {author} {\bibinfo {author} {\bibfnamefont {O.~G.}\ \bibnamefont
  {Benvenuto}}\ and\ \bibinfo {author} {\bibfnamefont {G.}~\bibnamefont
  {Lugones}},\ }\bibfield  {title} {\bibinfo {title} {{The properties of
  strange stars in the quark mass-density-dependent model}},\ }\href
  {https://doi.org/10.1142/S0218271898000048} {\bibfield  {journal} {\bibinfo
  {journal} {Int. J. Mod. Phys. D}\ }\textbf {\bibinfo {volume} {7}},\ \bibinfo
  {pages} {29} (\bibinfo {year} {1998})}\BibitemShut {NoStop}%
\bibitem [{\citenamefont {Lugones}\ and\ \citenamefont
  {Horvath}(2003)}]{Lugones:2002vd}%
  \BibitemOpen
  \bibfield  {author} {\bibinfo {author} {\bibfnamefont {G.}~\bibnamefont
  {Lugones}}\ and\ \bibinfo {author} {\bibfnamefont {J.~E.}\ \bibnamefont
  {Horvath}},\ }\bibfield  {title} {\bibinfo {title} {{Quark - diquark equation
  of state and compact star structure}},\ }\href
  {https://doi.org/10.1142/S0218271803002755} {\bibfield  {journal} {\bibinfo
  {journal} {Int. J. Mod. Phys. D}\ }\textbf {\bibinfo {volume} {12}},\
  \bibinfo {pages} {495} (\bibinfo {year} {2003})},\ \Eprint
  {https://arxiv.org/abs/astro-ph/0203069} {arXiv:astro-ph/0203069}
  \BibitemShut {NoStop}%
\bibitem [{\citenamefont {Peng}\ \emph
  {et~al.}(2000{\natexlab{a}})\citenamefont {Peng}, \citenamefont {Chiang},
  \citenamefont {Yang}, \citenamefont {Li},\ and\ \citenamefont
  {Liu}}]{Peng:1999gh}%
  \BibitemOpen
  \bibfield  {author} {\bibinfo {author} {\bibfnamefont {G.~X.}\ \bibnamefont
  {Peng}}, \bibinfo {author} {\bibfnamefont {H.~C.}\ \bibnamefont {Chiang}},
  \bibinfo {author} {\bibfnamefont {J.~J.}\ \bibnamefont {Yang}}, \bibinfo
  {author} {\bibfnamefont {L.}~\bibnamefont {Li}},\ and\ \bibinfo {author}
  {\bibfnamefont {B.}~\bibnamefont {Liu}},\ }\bibfield  {title} {\bibinfo
  {title} {{Mass formulas and thermodynamic treatment in the mass density
  dependent model of strange quark matter}},\ }\href
  {https://doi.org/10.1103/PhysRevC.61.015201} {\bibfield  {journal} {\bibinfo
  {journal} {Phys. Rev. C}\ }\textbf {\bibinfo {volume} {61}},\ \bibinfo
  {pages} {015201} (\bibinfo {year} {2000}{\natexlab{a}})},\ \Eprint
  {https://arxiv.org/abs/hep-ph/9911222} {arXiv:hep-ph/9911222} \BibitemShut
  {NoStop}%
\bibitem [{\citenamefont {Wang}(2000)}]{Wang:2000dc}%
  \BibitemOpen
  \bibfield  {author} {\bibinfo {author} {\bibfnamefont {P.}~\bibnamefont
  {Wang}},\ }\bibfield  {title} {\bibinfo {title} {{Strange matter in a
  selfconsistent quark mass density dependent model}},\ }\href
  {https://doi.org/10.1103/PhysRevC.62.015204} {\bibfield  {journal} {\bibinfo
  {journal} {Phys. Rev. C}\ }\textbf {\bibinfo {volume} {62}},\ \bibinfo
  {pages} {015204} (\bibinfo {year} {2000})}\BibitemShut {NoStop}%
\bibitem [{\citenamefont {Peng}\ \emph
  {et~al.}(2000{\natexlab{b}})\citenamefont {Peng}, \citenamefont {Chiang},\
  and\ \citenamefont {Ning}}]{Peng:2000ff}%
  \BibitemOpen
  \bibfield  {author} {\bibinfo {author} {\bibfnamefont {G.~X.}\ \bibnamefont
  {Peng}}, \bibinfo {author} {\bibfnamefont {H.~C.}\ \bibnamefont {Chiang}},\
  and\ \bibinfo {author} {\bibfnamefont {P.~Z.}\ \bibnamefont {Ning}},\
  }\bibfield  {title} {\bibinfo {title} {{Thermodynamics, strange quark matter,
  and strange stars}},\ }\href {https://doi.org/10.1103/PhysRevC.62.025801}
  {\bibfield  {journal} {\bibinfo  {journal} {Phys. Rev. C}\ }\textbf {\bibinfo
  {volume} {62}},\ \bibinfo {pages} {025801} (\bibinfo {year}
  {2000}{\natexlab{b}})},\ \Eprint {https://arxiv.org/abs/hep-ph/0003027}
  {arXiv:hep-ph/0003027} \BibitemShut {NoStop}%
\bibitem [{\citenamefont {Yin}\ and\ \citenamefont {Su}(2008)}]{Yin:2008me}%
  \BibitemOpen
  \bibfield  {author} {\bibinfo {author} {\bibfnamefont {S.-y.}\ \bibnamefont
  {Yin}}\ and\ \bibinfo {author} {\bibfnamefont {R.-K.}\ \bibnamefont {Su}},\
  }\bibfield  {title} {\bibinfo {title} {{Consistent thermodynamic treatment
  for a quark-mass density-dependent model}},\ }\href
  {https://doi.org/10.1103/PhysRevC.77.055204} {\bibfield  {journal} {\bibinfo
  {journal} {Phys. Rev. C}\ }\textbf {\bibinfo {volume} {77}},\ \bibinfo
  {pages} {055204} (\bibinfo {year} {2008})},\ \Eprint
  {https://arxiv.org/abs/0801.2813} {arXiv:0801.2813 [nucl-th]} \BibitemShut
  {NoStop}%
\bibitem [{\citenamefont {Xia}\ \emph {et~al.}(2014)\citenamefont {Xia},
  \citenamefont {Peng}, \citenamefont {Chen}, \citenamefont {Lu},\ and\
  \citenamefont {Xu}}]{Xia:2014zaa}%
  \BibitemOpen
  \bibfield  {author} {\bibinfo {author} {\bibfnamefont {C.~J.}\ \bibnamefont
  {Xia}}, \bibinfo {author} {\bibfnamefont {G.~X.}\ \bibnamefont {Peng}},
  \bibinfo {author} {\bibfnamefont {S.~W.}\ \bibnamefont {Chen}}, \bibinfo
  {author} {\bibfnamefont {Z.~Y.}\ \bibnamefont {Lu}},\ and\ \bibinfo {author}
  {\bibfnamefont {J.~F.}\ \bibnamefont {Xu}},\ }\bibfield  {title} {\bibinfo
  {title} {{Thermodynamic consistency, quark mass scaling, and properties of
  strange matter}},\ }\href {https://doi.org/10.1103/PhysRevD.89.105027}
  {\bibfield  {journal} {\bibinfo  {journal} {Phys. Rev. D}\ }\textbf {\bibinfo
  {volume} {89}},\ \bibinfo {pages} {105027} (\bibinfo {year} {2014})},\
  \Eprint {https://arxiv.org/abs/1405.3037} {arXiv:1405.3037 [hep-ph]}
  \BibitemShut {NoStop}%
\bibitem [{\citenamefont {Demorest}\ \emph {et~al.}(2010)\citenamefont
  {Demorest}, \citenamefont {Pennucci}, \citenamefont {Ransom}, \citenamefont
  {Roberts},\ and\ \citenamefont {Hessels}}]{Demorest:2010bx}%
  \BibitemOpen
  \bibfield  {author} {\bibinfo {author} {\bibfnamefont {P.}~\bibnamefont
  {Demorest}}, \bibinfo {author} {\bibfnamefont {T.}~\bibnamefont {Pennucci}},
  \bibinfo {author} {\bibfnamefont {S.}~\bibnamefont {Ransom}}, \bibinfo
  {author} {\bibfnamefont {M.}~\bibnamefont {Roberts}},\ and\ \bibinfo {author}
  {\bibfnamefont {J.}~\bibnamefont {Hessels}},\ }\bibfield  {title} {\bibinfo
  {title} {{Shapiro Delay Measurement of A Two Solar Mass Neutron Star}},\
  }\href {https://doi.org/10.1038/nature09466} {\bibfield  {journal} {\bibinfo
  {journal} {Nature}\ }\textbf {\bibinfo {volume} {467}},\ \bibinfo {pages}
  {1081} (\bibinfo {year} {2010})},\ \Eprint {https://arxiv.org/abs/1010.5788}
  {arXiv:1010.5788 [astro-ph.HE]} \BibitemShut {NoStop}%
\bibitem [{\citenamefont {Antoniadis}\ \emph {et~al.}(2013)\citenamefont
  {Antoniadis} \emph {et~al.}}]{Antoniadis:2013pzd}%
  \BibitemOpen
  \bibfield  {author} {\bibinfo {author} {\bibfnamefont {J.}~\bibnamefont
  {Antoniadis}} \emph {et~al.},\ }\bibfield  {title} {\bibinfo {title} {{A
  Massive Pulsar in a Compact Relativistic Binary}},\ }\href
  {https://doi.org/10.1126/science.1233232} {\bibfield  {journal} {\bibinfo
  {journal} {Science}\ }\textbf {\bibinfo {volume} {340}},\ \bibinfo {pages}
  {6131} (\bibinfo {year} {2013})},\ \Eprint {https://arxiv.org/abs/1304.6875}
  {arXiv:1304.6875 [astro-ph.HE]} \BibitemShut {NoStop}%
\bibitem [{\citenamefont {Cromartie}\ \emph {et~al.}(2019)\citenamefont
  {Cromartie} \emph {et~al.}}]{NANOGrav:2019jur}%
  \BibitemOpen
  \bibfield  {author} {\bibinfo {author} {\bibfnamefont {H.~T.}\ \bibnamefont
  {Cromartie}} \emph {et~al.} (\bibinfo {collaboration} {NANOGrav}),\
  }\bibfield  {title} {\bibinfo {title} {{Relativistic Shapiro delay
  measurements of an extremely massive millisecond pulsar}},\ }\href
  {https://doi.org/10.1038/s41550-019-0880-2} {\bibfield  {journal} {\bibinfo
  {journal} {Nature Astron.}\ }\textbf {\bibinfo {volume} {4}},\ \bibinfo
  {pages} {72} (\bibinfo {year} {2019})},\ \Eprint
  {https://arxiv.org/abs/1904.06759} {arXiv:1904.06759 [astro-ph.HE]}
  \BibitemShut {NoStop}%
\bibitem [{\citenamefont {Riley}\ \emph {et~al.}(2021)\citenamefont {Riley}
  \emph {et~al.}}]{Riley:2021pdl}%
  \BibitemOpen
  \bibfield  {author} {\bibinfo {author} {\bibfnamefont {T.~E.}\ \bibnamefont
  {Riley}} \emph {et~al.},\ }\bibfield  {title} {\bibinfo {title} {{A NICER
  View of the Massive Pulsar PSR J0740+6620 Informed by Radio Timing and
  XMM-Newton Spectroscopy}},\ }\href {https://doi.org/10.3847/2041-8213/ac0a81}
  {\bibfield  {journal} {\bibinfo  {journal} {Astrophys. J. Lett.}\ }\textbf
  {\bibinfo {volume} {918}},\ \bibinfo {pages} {L27} (\bibinfo {year}
  {2021})},\ \Eprint {https://arxiv.org/abs/2105.06980} {arXiv:2105.06980
  [astro-ph.HE]} \BibitemShut {NoStop}%
\bibitem [{\citenamefont {Miller}\ \emph {et~al.}(2021)\citenamefont {Miller}
  \emph {et~al.}}]{Miller:2021qha}%
  \BibitemOpen
  \bibfield  {author} {\bibinfo {author} {\bibfnamefont {M.~C.}\ \bibnamefont
  {Miller}} \emph {et~al.},\ }\bibfield  {title} {\bibinfo {title} {{The Radius
  of PSR J0740+6620 from NICER and XMM-Newton Data}},\ }\href
  {https://doi.org/10.3847/2041-8213/ac089b} {\bibfield  {journal} {\bibinfo
  {journal} {Astrophys. J. Lett.}\ }\textbf {\bibinfo {volume} {918}},\
  \bibinfo {pages} {L28} (\bibinfo {year} {2021})},\ \Eprint
  {https://arxiv.org/abs/2105.06979} {arXiv:2105.06979 [astro-ph.HE]}
  \BibitemShut {NoStop}%
\bibitem [{\citenamefont {Rischke}\ \emph {et~al.}(1991)\citenamefont
  {Rischke}, \citenamefont {Gorenstein}, \citenamefont {Stoecker},\ and\
  \citenamefont {Greiner}}]{Rischke:1991ke}%
  \BibitemOpen
  \bibfield  {author} {\bibinfo {author} {\bibfnamefont {D.~H.}\ \bibnamefont
  {Rischke}}, \bibinfo {author} {\bibfnamefont {M.~I.}\ \bibnamefont
  {Gorenstein}}, \bibinfo {author} {\bibfnamefont {H.}~\bibnamefont
  {Stoecker}},\ and\ \bibinfo {author} {\bibfnamefont {W.}~\bibnamefont
  {Greiner}},\ }\bibfield  {title} {\bibinfo {title} {{Excluded volume effect
  for the nuclear matter equation of state}},\ }\href
  {https://doi.org/10.1007/BF01548574} {\bibfield  {journal} {\bibinfo
  {journal} {Z. Phys. C}\ }\textbf {\bibinfo {volume} {51}},\ \bibinfo {pages}
  {485} (\bibinfo {year} {1991})}\BibitemShut {NoStop}%
\bibitem [{\citenamefont {Yen}\ \emph {et~al.}(1997)\citenamefont {Yen},
  \citenamefont {Gorenstein}, \citenamefont {Greiner},\ and\ \citenamefont
  {Yang}}]{Yen:1997rv}%
  \BibitemOpen
  \bibfield  {author} {\bibinfo {author} {\bibfnamefont {G.~D.}\ \bibnamefont
  {Yen}}, \bibinfo {author} {\bibfnamefont {M.~I.}\ \bibnamefont {Gorenstein}},
  \bibinfo {author} {\bibfnamefont {W.}~\bibnamefont {Greiner}},\ and\ \bibinfo
  {author} {\bibfnamefont {S.-N.}\ \bibnamefont {Yang}},\ }\bibfield  {title}
  {\bibinfo {title} {{Excluded volume hadron gas model for particle number
  ratios in A+A collisions}},\ }\href
  {https://doi.org/10.1103/PhysRevC.56.2210} {\bibfield  {journal} {\bibinfo
  {journal} {Phys. Rev. C}\ }\textbf {\bibinfo {volume} {56}},\ \bibinfo
  {pages} {2210} (\bibinfo {year} {1997})},\ \Eprint
  {https://arxiv.org/abs/nucl-th/9711062} {arXiv:nucl-th/9711062} \BibitemShut
  {NoStop}%
\bibitem [{\citenamefont {Steinheimer}\ \emph {et~al.}(2011)\citenamefont
  {Steinheimer}, \citenamefont {Schramm},\ and\ \citenamefont
  {Stocker}}]{Steinheimer:2010ib}%
  \BibitemOpen
  \bibfield  {author} {\bibinfo {author} {\bibfnamefont {J.}~\bibnamefont
  {Steinheimer}}, \bibinfo {author} {\bibfnamefont {S.}~\bibnamefont
  {Schramm}},\ and\ \bibinfo {author} {\bibfnamefont {H.}~\bibnamefont
  {Stocker}},\ }\bibfield  {title} {\bibinfo {title} {{An Effective chiral
  Hadron-Quark Equation of State}},\ }\href
  {https://doi.org/10.1088/0954-3899/38/3/035001} {\bibfield  {journal}
  {\bibinfo  {journal} {J. Phys. G}\ }\textbf {\bibinfo {volume} {38}},\
  \bibinfo {pages} {035001} (\bibinfo {year} {2011})},\ \Eprint
  {https://arxiv.org/abs/1009.5239} {arXiv:1009.5239 [hep-ph]} \BibitemShut
  {NoStop}%
\bibitem [{\citenamefont {Callen}(1985)}]{Callen}%
  \BibitemOpen
  \bibfield  {author} {\bibinfo {author} {\bibfnamefont {H.}~\bibnamefont
  {Callen}},\ }\href@noop {} {\emph {\bibinfo {title} {Thermodynamics and an
  Introduction to Thermostatistics}}}\ (\bibinfo  {publisher} {Wiley},\
  \bibinfo {year} {1985})\BibitemShut {NoStop}%
\bibitem [{\citenamefont {Riley}\ \emph {et~al.}(2019)\citenamefont {Riley}
  \emph {et~al.}}]{Riley:2019yda}%
  \BibitemOpen
  \bibfield  {author} {\bibinfo {author} {\bibfnamefont {T.~E.}\ \bibnamefont
  {Riley}} \emph {et~al.},\ }\bibfield  {title} {\bibinfo {title} {{A NICER
  View of PSR J0030+0451: Millisecond Pulsar Parameter Estimation}},\ }\href
  {https://doi.org/10.3847/2041-8213/ab481c} {\bibfield  {journal} {\bibinfo
  {journal} {Astrophys. J. Lett.}\ }\textbf {\bibinfo {volume} {887}},\
  \bibinfo {pages} {L21} (\bibinfo {year} {2019})},\ \Eprint
  {https://arxiv.org/abs/1912.05702} {arXiv:1912.05702 [astro-ph.HE]}
  \BibitemShut {NoStop}%
\bibitem [{\citenamefont {Miller}\ \emph {et~al.}(2019)\citenamefont {Miller}
  \emph {et~al.}}]{Miller:2019cac}%
  \BibitemOpen
  \bibfield  {author} {\bibinfo {author} {\bibfnamefont {M.}~\bibnamefont
  {Miller}} \emph {et~al.},\ }\bibfield  {title} {\bibinfo {title} {{PSR
  J0030+0451 Mass and Radius from NICER Data and Implications for the
  Properties of Neutron Star Matter}},\ }\href
  {https://doi.org/10.3847/2041-8213/ab50c5} {\bibfield  {journal} {\bibinfo
  {journal} {Astrophys. J. Lett.}\ }\textbf {\bibinfo {volume} {887}},\
  \bibinfo {pages} {L24} (\bibinfo {year} {2019})},\ \Eprint
  {https://arxiv.org/abs/1912.05705} {arXiv:1912.05705 [astro-ph.HE]}
  \BibitemShut {NoStop}%
\bibitem [{\citenamefont {Abbott}\ \emph {et~al.}(2018)\citenamefont {Abbott}
  \emph {et~al.}}]{Abbott:2018exr}%
  \BibitemOpen
  \bibfield  {author} {\bibinfo {author} {\bibfnamefont {B.}~\bibnamefont
  {Abbott}} \emph {et~al.} (\bibinfo {collaboration} {LIGO Scientific,
  Virgo}),\ }\bibfield  {title} {\bibinfo {title} {{GW170817: Measurements of
  neutron star radii and equation of state}},\ }\href
  {https://doi.org/10.1103/PhysRevLett.121.161101} {\bibfield  {journal}
  {\bibinfo  {journal} {Phys. Rev. Lett.}\ }\textbf {\bibinfo {volume} {121}},\
  \bibinfo {pages} {161101} (\bibinfo {year} {2018})},\ \Eprint
  {https://arxiv.org/abs/1805.11581} {arXiv:1805.11581 [gr-qc]} \BibitemShut
  {NoStop}%
\bibitem [{\citenamefont {M\"uther}\ \emph {et~al.}(1987)\citenamefont
  {M\"uther}, \citenamefont {Prakash},\ and\ \citenamefont
  {Ainsworth}}]{Muther:1987xaa}%
  \BibitemOpen
  \bibfield  {author} {\bibinfo {author} {\bibfnamefont {H.}~\bibnamefont
  {M\"uther}}, \bibinfo {author} {\bibfnamefont {M.}~\bibnamefont {Prakash}},\
  and\ \bibinfo {author} {\bibfnamefont {T.~L.}\ \bibnamefont {Ainsworth}},\
  }\bibfield  {title} {\bibinfo {title} {{The nuclear symmetry energy in
  relativistic Brueckner-Hartree-Fock calculations}},\ }\href
  {https://doi.org/10.1016/0370-2693(87)91611-X} {\bibfield  {journal}
  {\bibinfo  {journal} {Phys. Lett. B}\ }\textbf {\bibinfo {volume} {199}},\
  \bibinfo {pages} {469} (\bibinfo {year} {1987})}\BibitemShut {NoStop}%
\bibitem [{\citenamefont {{Doroshenko}}\ \emph {et~al.}(2022)\citenamefont
  {{Doroshenko}}, \citenamefont {{Suleimanov}}, \citenamefont
  {{P{\"u}hlhofer}},\ and\ \citenamefont {{Santangelo}}}]{HESS2022}%
  \BibitemOpen
  \bibfield  {author} {\bibinfo {author} {\bibfnamefont {V.}~\bibnamefont
  {{Doroshenko}}}, \bibinfo {author} {\bibfnamefont {V.}~\bibnamefont
  {{Suleimanov}}}, \bibinfo {author} {\bibfnamefont {G.}~\bibnamefont
  {{P{\"u}hlhofer}}},\ and\ \bibinfo {author} {\bibfnamefont {A.}~\bibnamefont
  {{Santangelo}}},\ }\bibfield  {title} {\bibinfo {title} {{A strangely light
  neutron star within a supernova remnant}},\ }\href
  {https://doi.org/10.1038/s41550-022-01800-1} {\bibfield  {journal} {\bibinfo
  {journal} {Nature Astronomy}\ }\textbf {\bibinfo {volume} {6}},\ \bibinfo
  {pages} {1444} (\bibinfo {year} {2022})}\BibitemShut {NoStop}%
\end{thebibliography}%
\end{document}